\documentclass[a4paper,aps,prd,10pt,preprintnumbers,twocolumn,superscriptaddress,nofootinbib,amsmath,amssymb]{revtex4-1}
\usepackage{graphicx}
\usepackage[utf8]{inputenc}
\usepackage[T1]{fontenc}
\usepackage{cmap}
\usepackage{float}

\DeclareMathAlphabet{\pazocal}{OMS}{zplm}{m}{n}

\begin{document}
\title{Rotating Bardeen black hole surrounded by perfect fluid dark matter as a particle accelerator  }
\author{Qi-Quan Li}
\author{Yu Zhang} \email{zhangyu_128@126.com (Corresponding author)}
\author{Qian Li}
\author{Qi Sun}
\affiliation{Faculty of Science, Kunming University of Science and Technology, Kunming, Yunnan 650500, China}
\begin{abstract}
We study the event horizon of a rotating Bardeen black hole surrounded by perfect fluid dark matter and the black hole as a particle accelerator. The black hole is represented by four parameters: mass $M$, rotation parameter $a$, dark matter parameter $\alpha$ and magnetic charge $g$. It is interesting that when we determine the values of magnetic charge $g$ and dark matter parameters $\alpha$ we can get a critical rotation parameter $a_E$ and then we get a contour plane with $\Delta$= 0 taking three parameters as coordinates. We also derive the effective potential of the particle and the centre-of-mass (CM) energy of the two particles outside the black hole by using the motion equation of the particle in the equatorial plane of the black hole. We find that the CM energy depends not only on the rotation parameter $a$, but also on the parameters $g$ and $\alpha$. We discuss the CM energy for two particles colliding at the black hole horizon in the extreme and non-extreme cases, respectively. It is found that the CM energy can be arbitrarily high when the angular momentum of one of the two particles is the critical angular momentum under the background of extreme black holes and there is no such result for non-extreme black holes, because the particles do not reach the black hole horizon when the angular momentum of the particles is critical angular momentum. Therefore, we prove that the rotating Bardeen black hole surrounded by perfect fluid dark matter can be used as a particle accelerator.
\end{abstract}
\maketitle

\section{Introduction}
In April 2019, the Event Horizon Telescope collaboration released the first image of a black hole shadow in Virgo A.
galaxy (M87), which lies 55 million light-years away from Earth \cite{EventHorizonTelescope:2019dse}. Just three years later, the collaboration officially published the first picture of the black hole shadow of Sgr A* located in the center of the Milky way \cite{EventHorizonTelescope:2022xnr}. The release of black hole shadows has given us an intuitive understanding of this mysterious celestial body in the universe. However, there are still many problems in the theory of black holes, one of which is the singularity of black holes. In 1970, Hawking and Penrose proposed that when a spacetime satisfies the four physical hypotheses \cite{Hawking:1970zqf}, it must be non-space-like and incomplete (i.e., there is a singularity). This means that when the mass of a star is greater than the Oppenheimer limit, it will collapse infinitely to form a black hole, and its mass will be compressed at the singularity of the black hole. The black hole singularity has a series of singular properties where the density and curvature are infinite, and the laws of physics as we know them break down at the singularity.

In the real physical world, it is of great interest whether there is a singularity and whether there is a theory that can avoid the occurrence of black hole singularity. There is no experimental confirmation of the existence of black hole singularities, but theoretically, a black hole singularity can be avoided. In 1968, Bardeen was the first to obtain a black hole solution with rules and no singularities everywhere \cite{Bardeen:1968}. Borde discussed and analyzed Bardeen black hole model in Refs.\cite{Borde:1994ai,Borde:1996df} and his study of black hole singularities further showed that in a large class of black holes, singularity can be avoided by topological changes \cite{Borde:1996df}.  Ay\'on-Beato and Garc\'ia then found that in the framework of the standard General Relativity one can find singularity-free
solutions of the Einstein field equations coupled to a suitable nonlinear electrodynamics, which is the charged version of the Bardeen black hole solution \cite{Ayon-Beato:1998hmi}. Bambi and Modesto proposed rotating Bardeen black hole \cite{Bambi:2013ufa}. In recent, the metric of non-rotating and rotating Bardeen black hole surrounded by perfect fluid dark matter have been proposed and painted the shapes of the Ergospheres \cite{Zhang:2020mxi}. Of course, there are other regular black holes, such as Hayward black hole \cite{Hayward:2006} and Berej-Matyjasek-Trynieki-Wornowicz black hole \cite{Berej:2006cc}.

In order to study physical problems such as the origin of the universe and the composition of matter, we built particle accelerators and particle colliders to accelerate
particle collisions but now we can only get particle collision
energy less than 10Tev, which is a huge gap with the energy corresponding to the physical theory that we need to verify and develop, such as the exploration of Planck-scale physics. Penrose indicated that black holes can provide a source of energy for surrounding matter \cite{Penrose:1971uk}. In 2009, Banados, Silk and West(BSW) \cite{Banados:2009pr} found that under the background of extreme Kerr black holes, when two particles of the same mass falling from infinity, if one of the particles has critical angular momentum, its center-of-mass (CM) energy is arbitrarily high when the two particles collide at the event horizon of the black hole. In Ref.\cite{Mao:2010di}, it was found that if a particle involved in the collision under the background
of charged non-rotating Kaluza-Klein black hole has a
critical charge, its CM energy will diverge at the horizon of the black hole, proving that non-extreme black
hole can also be used as a particle accelerator. Patil and Joshi studied CM energy around naked singularities \cite{Patil:2011aa,Patil:2012fu}, and they succeeded in obtaining a large amount of energy near the singularity. Since then, many different black holes have been demonstrated as particle accelerators \cite{Shaymatov:2020bso,Wei:2010vca,Li:2010ej,Zaslavskii:2010aw,Vrba:2019vqh,AhmedRizwan:2020sza,Babar:2021nst}.

In this paper, we mainly study the properties of particle collisions in the background of rotating Bardeen black hole surrounded by perfect fluid dark matter, and the basic structure of this paper is as follows. In Sec.~\ref{sec:2}, we review rotating Bardeen black hole surrounded by perfect fluid dark matter and analyze the influence of dark matter parameter $\alpha$ and rotation parameter $a$ and magnetic charge $g$ on the horizon structure of black hole. In Sec.~\ref{sec:3}, we calculate the equation of motion and effective potential of the particle on the equatorial plane of the black hole and discuss the range of angular momentum of the incident particle. In Sec.~\ref{sec:4}, we derive the CM energies of two particles, and discuss the CM energies of particles at the event horizon of extreme and non-extreme black holes. In Sec.~\ref{sec:5}, we summarize the content of the article.

\begin{figure*}
	\begin{tabular}{c c}
		\includegraphics[scale=0.5]{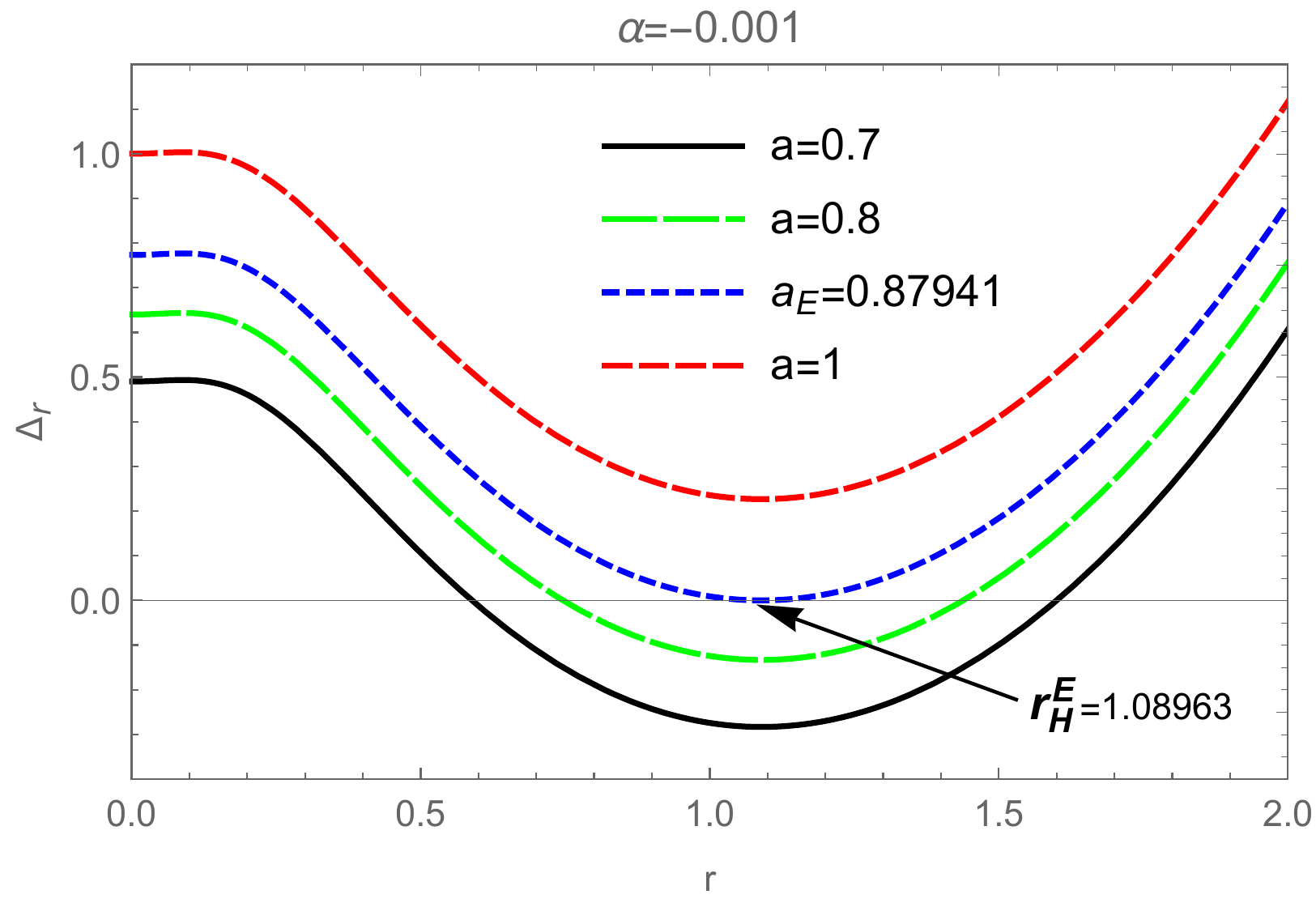}\hspace{-0.2cm}
		&\includegraphics[scale=0.5]{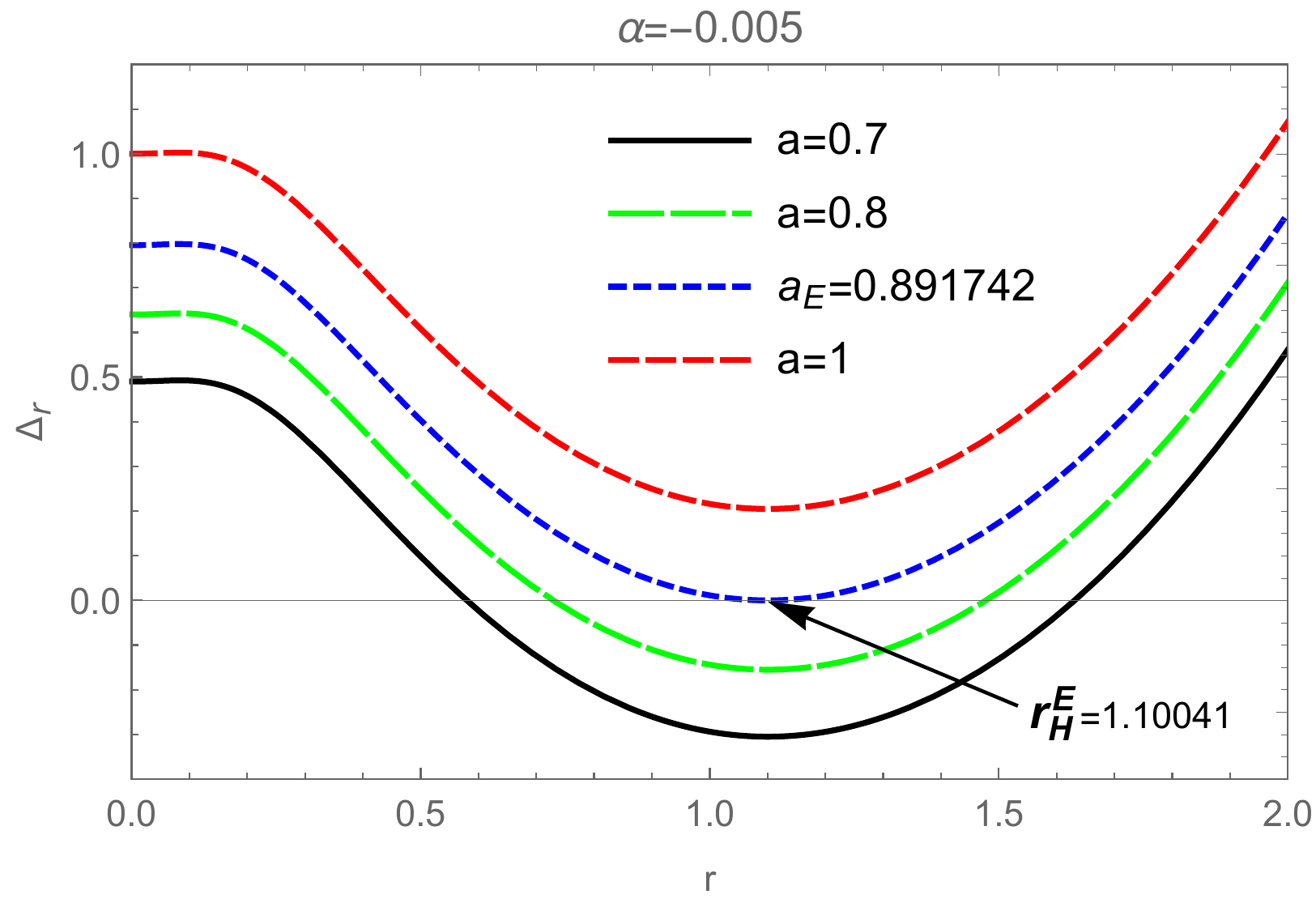}\\
		\includegraphics[scale=0.5]{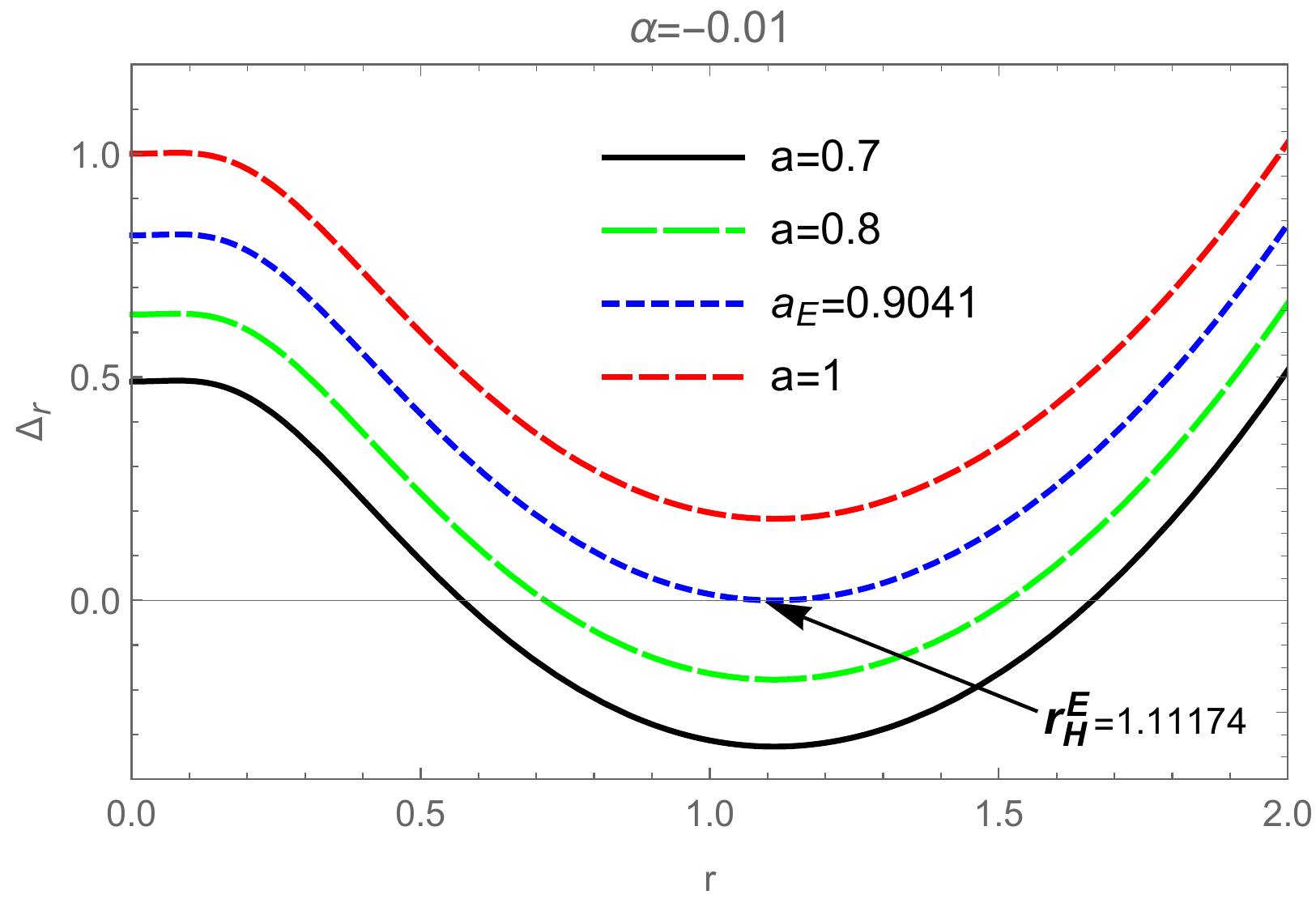}\hspace{-0.2cm}
		&\includegraphics[scale=0.5]{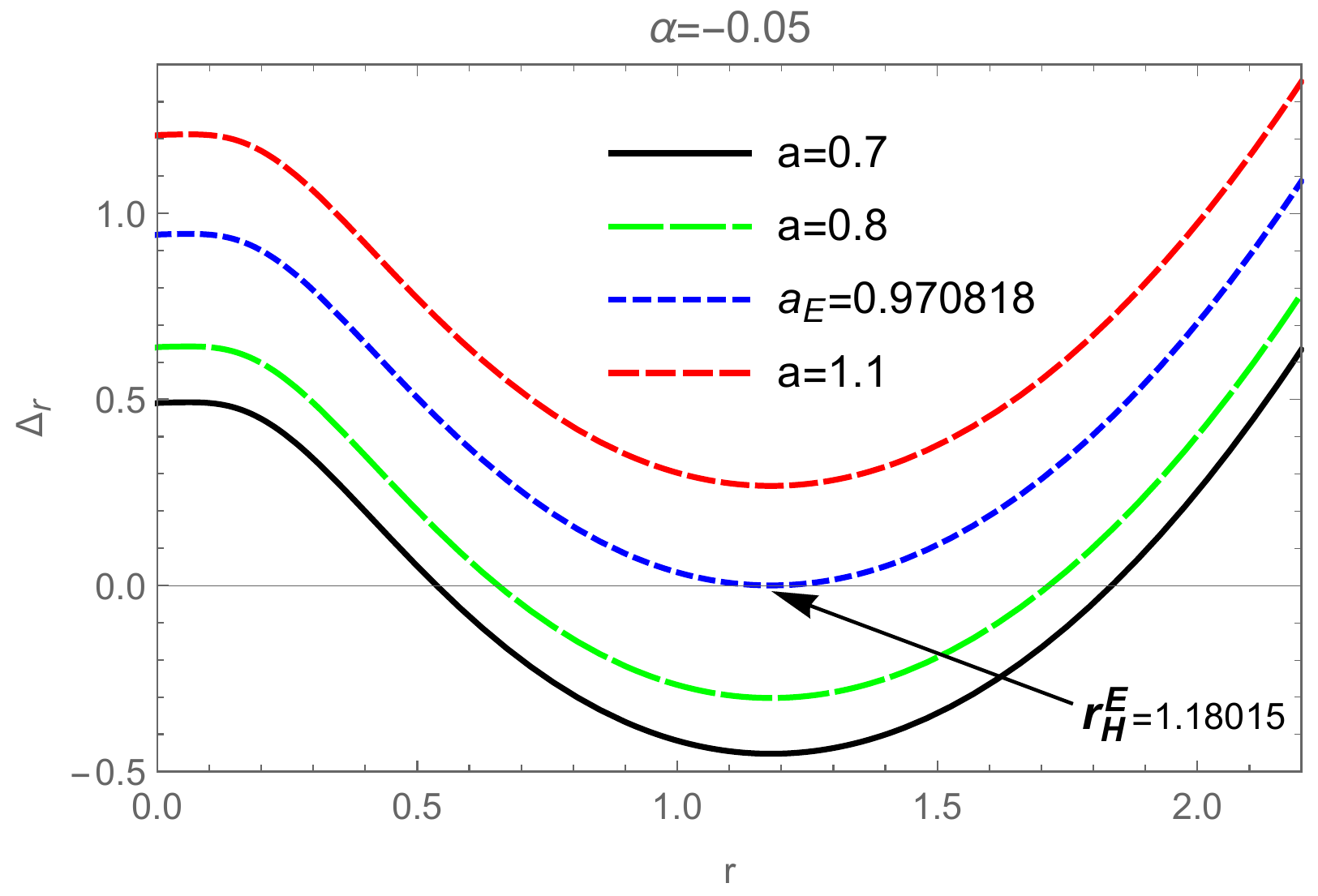}
	\end{tabular}
	\caption{The figure shows $\Delta$ vs $r$ for fixed values of $g=0.3$ and $M = 1$. Case  $a=a_E$ corresponds to an extremal black hole.}\label{fig:horizon1}
\end{figure*}

\begin{figure*}
	\begin{tabular}{c c}
		\includegraphics[scale=0.5]{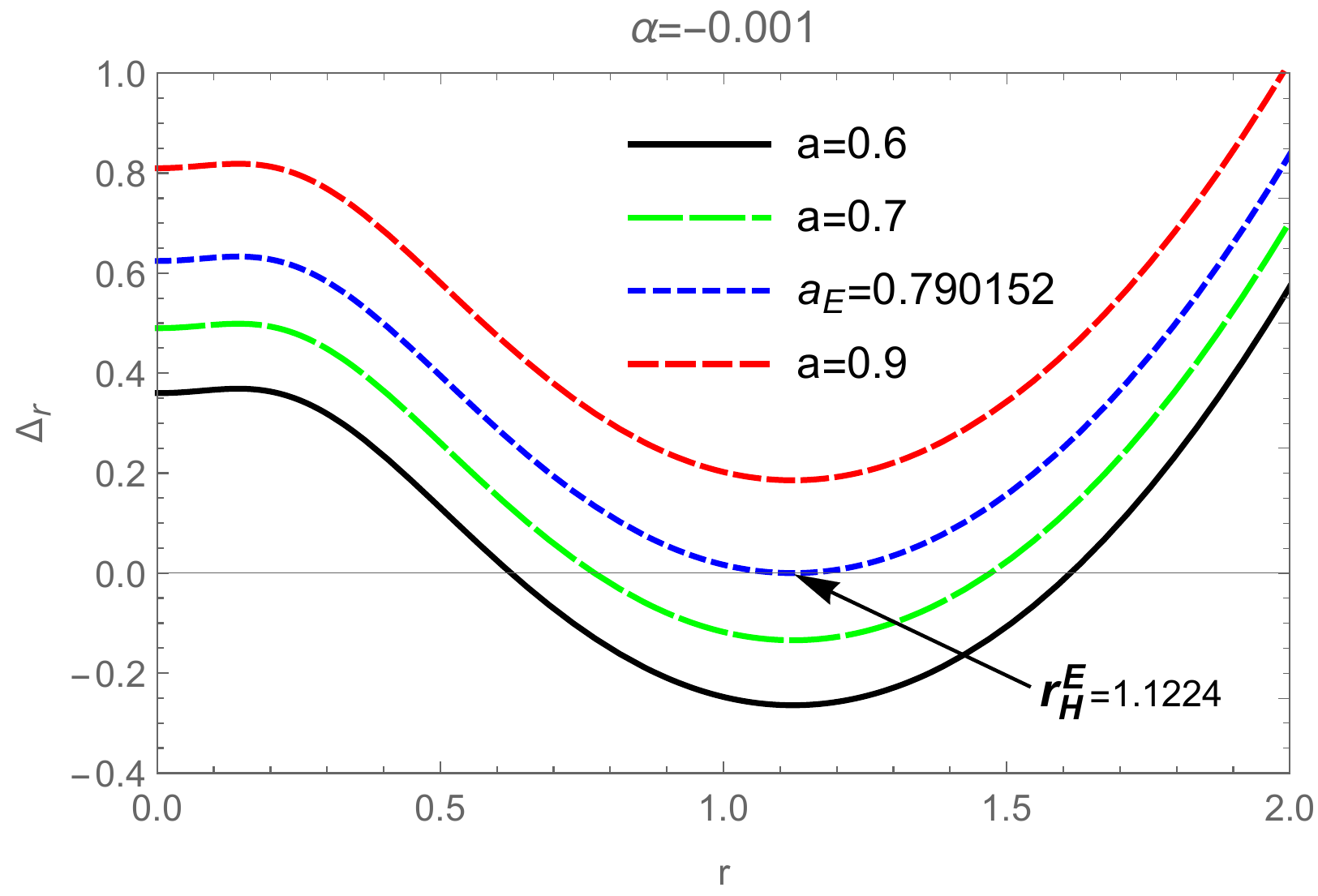}\hspace{-0.2cm}
		&\includegraphics[scale=0.5]{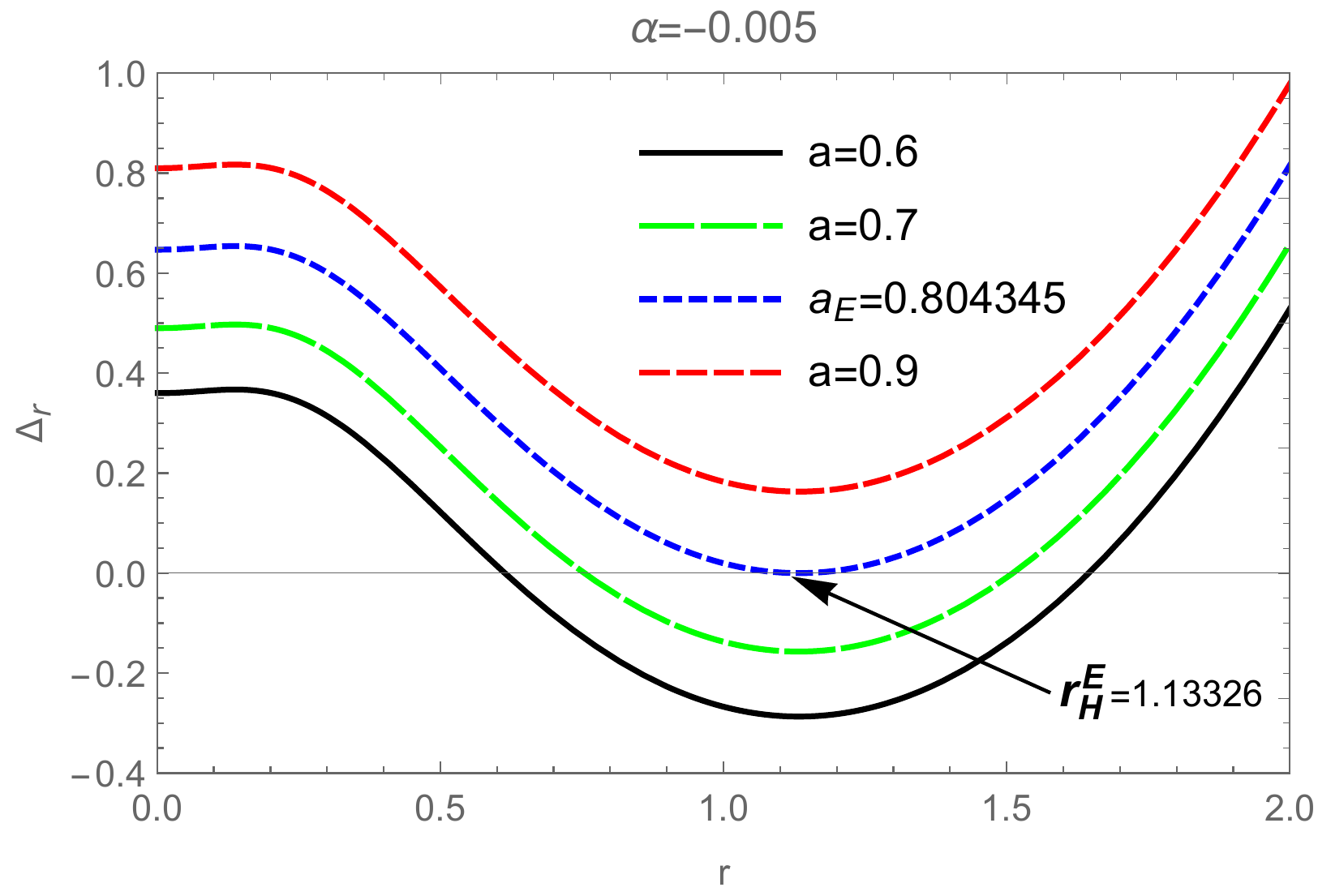}\\
		\includegraphics[scale=0.5]{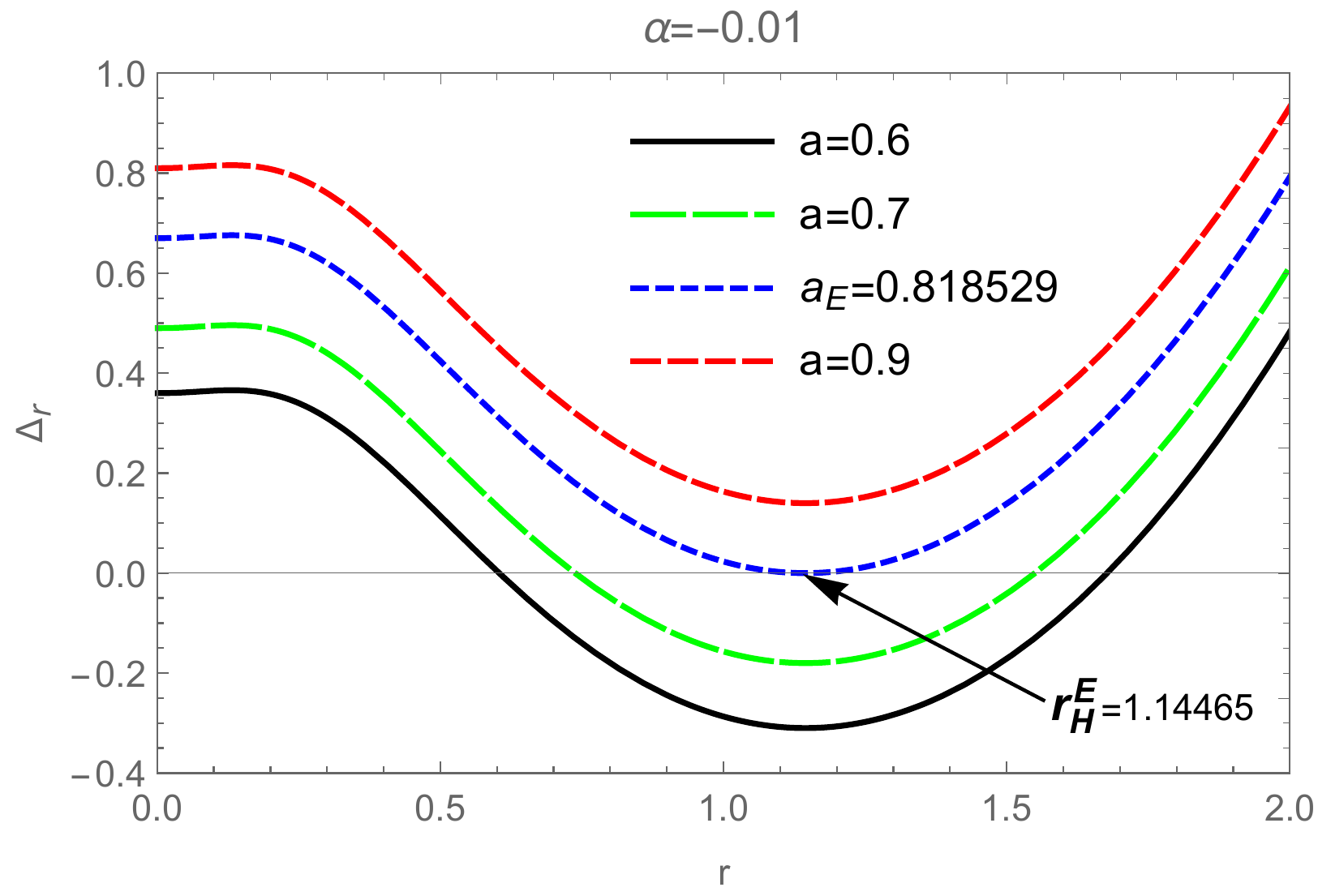}\hspace{-0.2cm}
		&\includegraphics[scale=0.5]{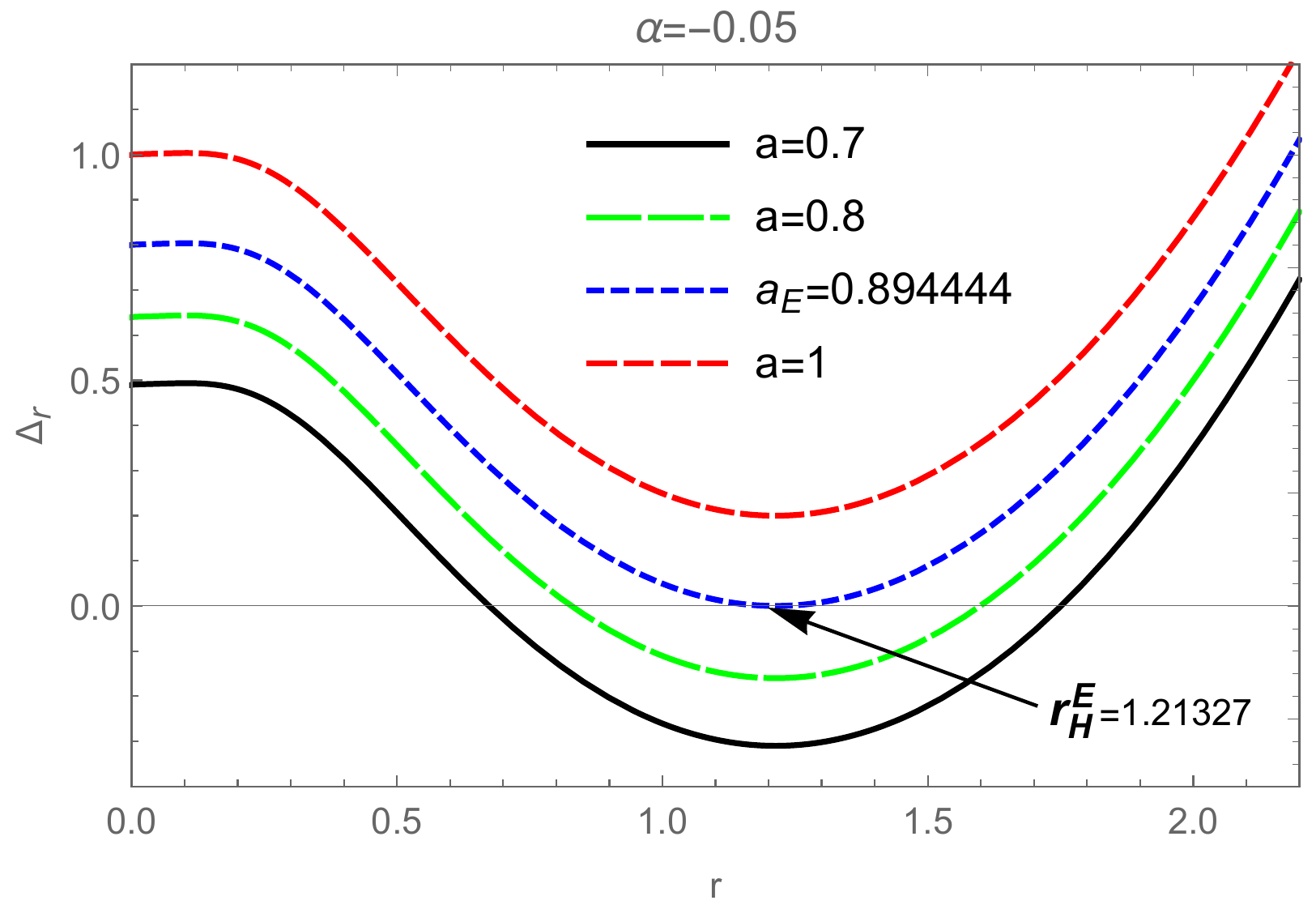}
	\end{tabular}
	\caption{The figure shows $\Delta$ vs $r$ for fixed values of $g=0.5$ and $M = 1$. Case  $a=a_E$ corresponds to an extremal black hole.}\label{fig:horizon2}
\end{figure*}

\begin{figure*}
	\begin{tabular}{c c c c}
		\includegraphics[scale=0.1]{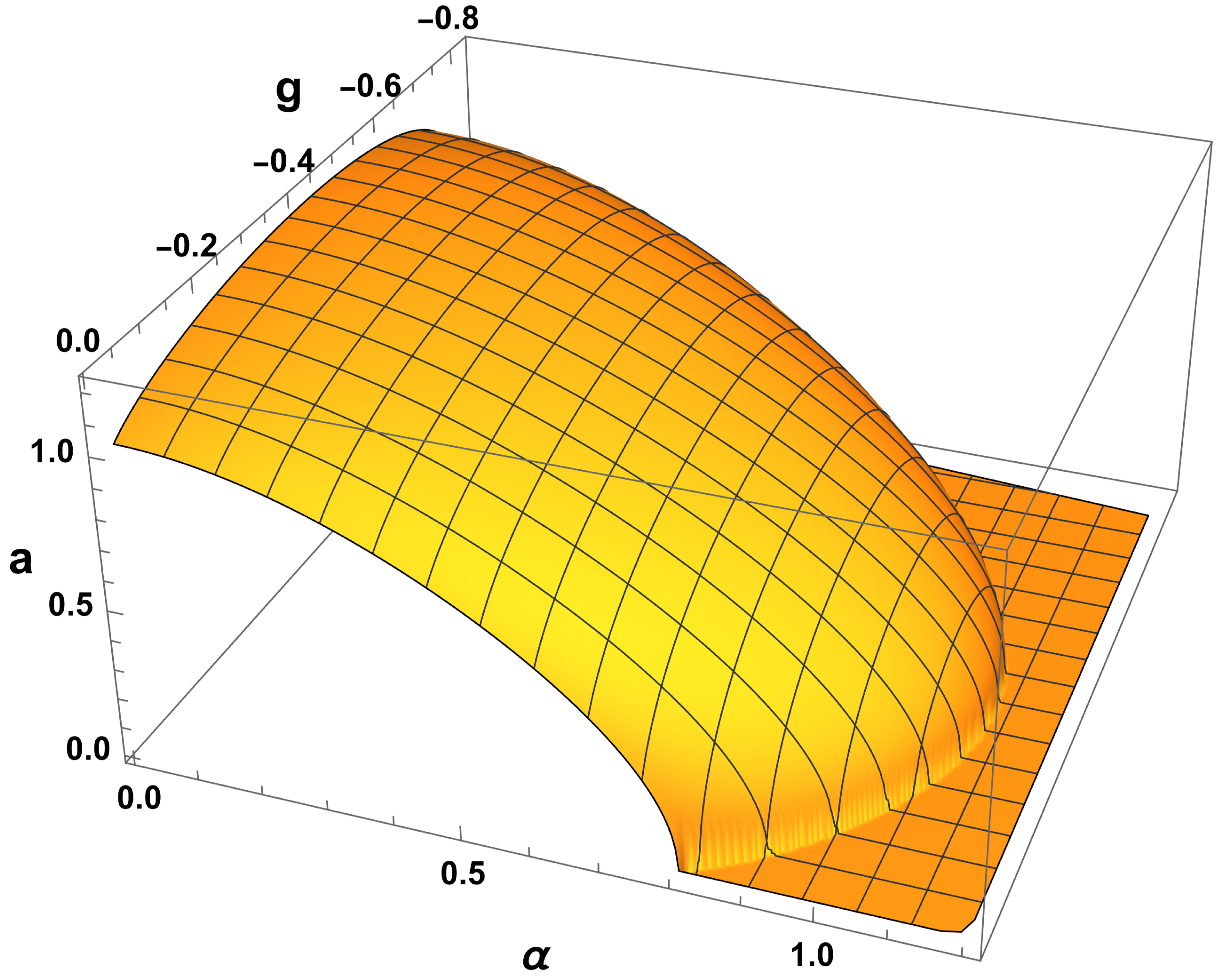}\hspace{-0cm}
		\includegraphics[scale=0.1]{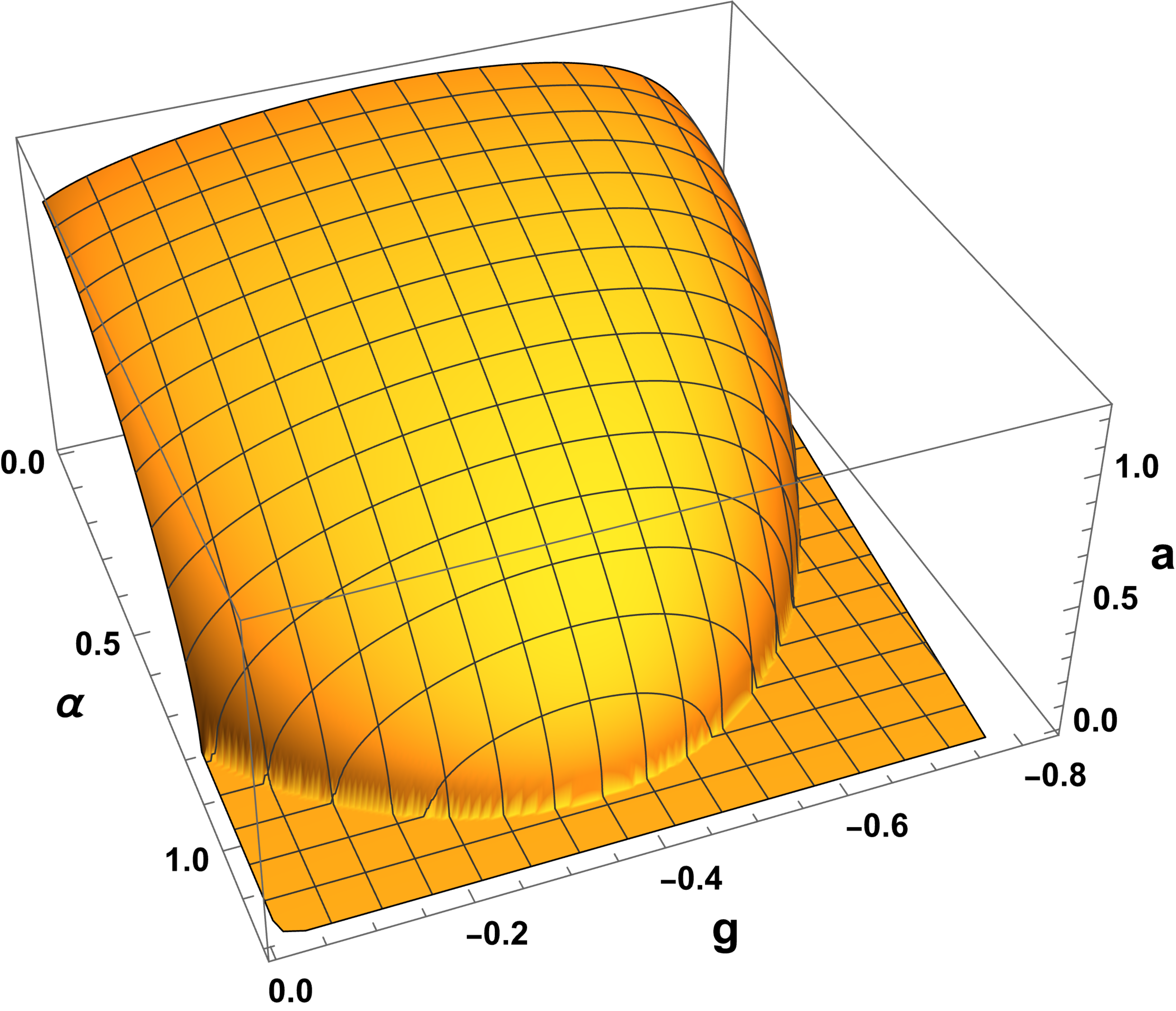}\\
		\includegraphics[scale=0.1]{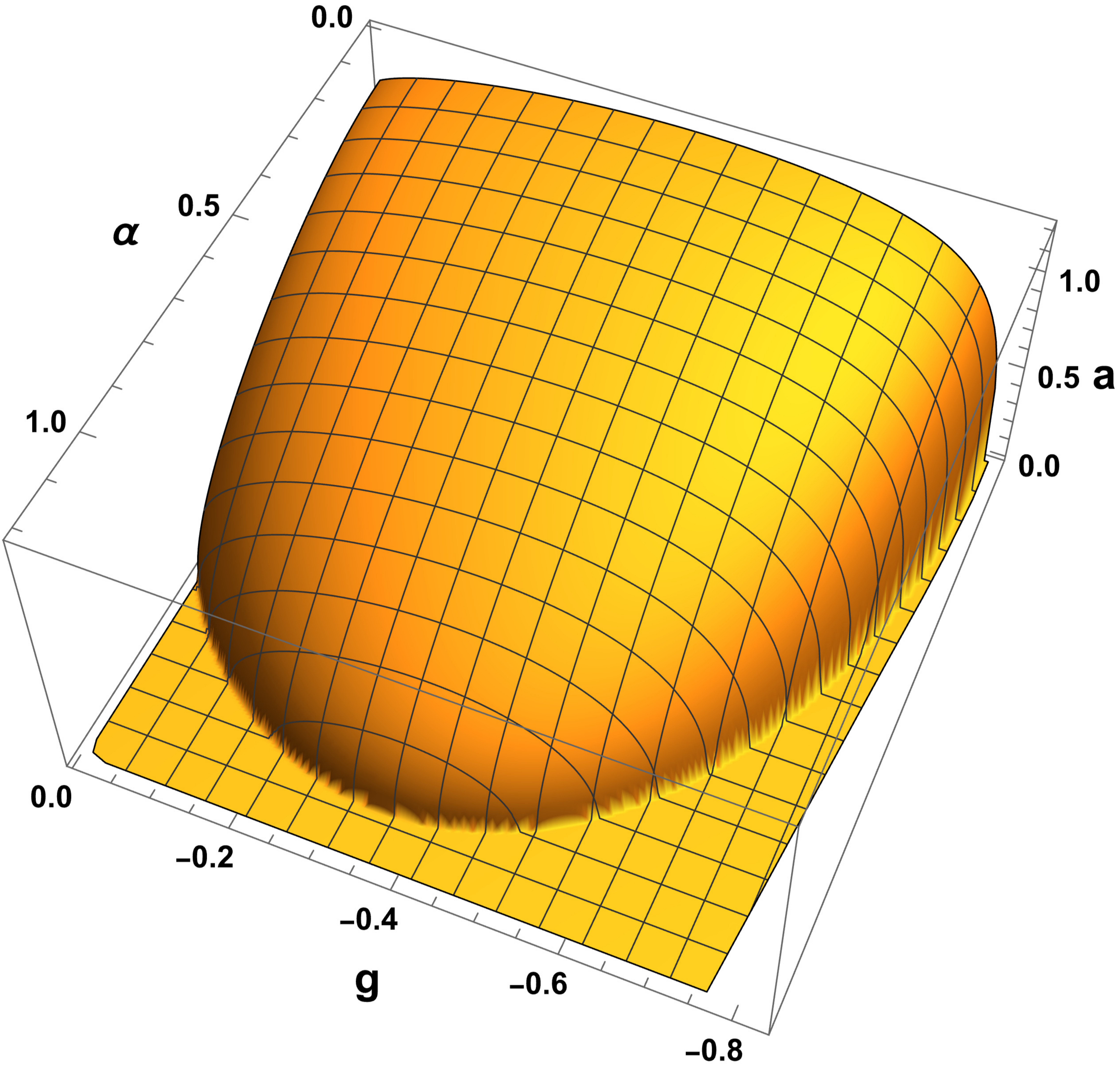}\hspace{-0cm}
		\includegraphics[scale=0.1]{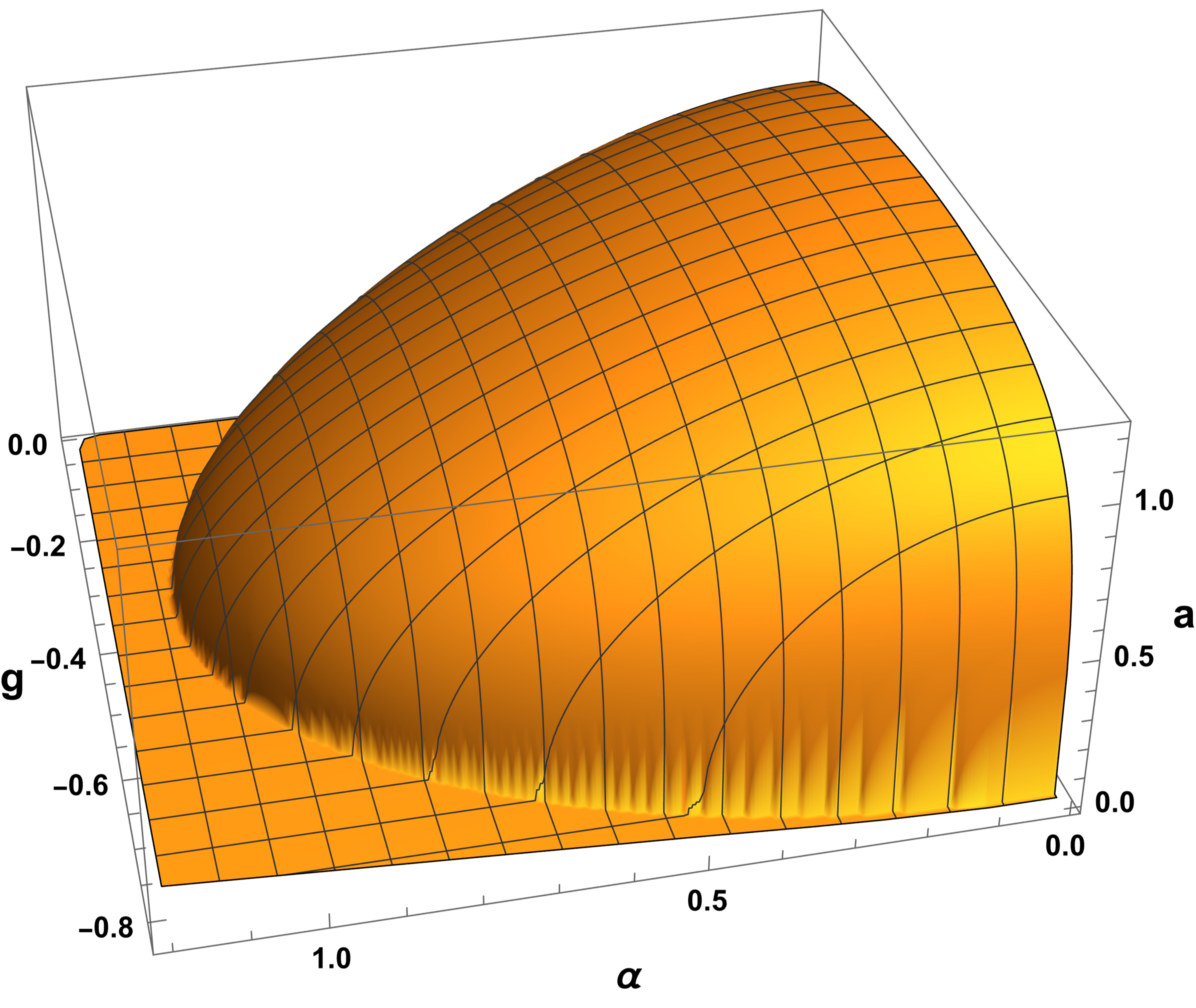}\\
	\end{tabular}
	\caption{The figure shows the contour plane of $\Delta_r=0$, and the three coordinates are magnetic charge parameter $g$, dark matter parameter $\alpha$  and rotation parameter $a$, respectively.}\label{fig:horizon3}
\end{figure*}

\section{ROTATING BARDEEN BLACK HOLE SURROUNDED BY PERFECT FLUID DARK MATTER}
\label{sec:2}
In this section, let us review the work of Zhang et al \cite{Zhang:2020mxi}. In the Bardeen model, considering the coupling of gravity and non-linear electromagnetic field, the corresponding Einstein-Maxwell equations should be modified as

\begin{equation}
	G_{\mu}^{\nu}=2\left(\frac{\partial \mathcal{L}(F)}{\partial F}
	F_{\mu \lambda} F^{\nu \lambda}-\delta_{\mu}^{\nu} \mathcal{L}\right)+8 \pi T_{\mu}^{\nu},
\end{equation}
\begin{equation}
	\nabla_{\mu}\left(\frac{\partial \mathcal{L}(F)}{\partial F} F^{\nu \mu}\right)=0 ,
\end{equation}
\begin{equation}
	\begin{array}{c}
		\nabla_{\mu}\left(* F^{\nu \mu}\right)=0.
	\end{array}
\end{equation}

Here, $ F_{\mu \nu}=2 \nabla_{[\mu} A_{\nu]}$  and $ \mathcal{L} $ is a function of  $ F \equiv \frac{1}{4} F_{\mu \nu} F^{\mu \nu}$, given by \cite{Ayon-Beato:2000mjt}
\begin{equation}
	\begin{array}{c}
		\mathcal{L}(F)=\frac{3 M}{|g|^{3}}\left(\frac{\sqrt{2 g^{2} F}}{1+\sqrt{2 g^{2} F}}\right)^{\frac{5}{2}},
	\end{array}
\end{equation}

where $g$ and $M$ are the magnetic charge and mass, respectively. Considering a black hole surrounded by perfect fluid dark matter, the energy-momentum tensor can be written as \cite{Kiselev:2002dx,Li:2012zx}

\begin{equation}
	\begin{array}{c}
		T_{\nu}^{\mu}=\operatorname{diag}\left(-\epsilon, p_{r}, p_{\theta}, p_{\phi}\right),
	\end{array}
\end{equation}
where
\begin{equation}
	\begin{array}{c}
		-\epsilon=p_{r}=\frac{\alpha}{8 \pi r^{3}} \quad \text { and } \quad p_{\theta}=p_{\phi}=-\frac{\alpha}{16 \pi r^{3}} .
	\end{array}
\end{equation}

The black hole solution of the Bardeen black hole surrounded by perfect fluid dark matter can be obtained by solving the Einstein-Maxwell equations and energy momentum tensor \cite{Zhang:2020mxi}

\begin{equation}
	\begin{aligned}
		\begin{array}{c}
			ds^2=-f(r) \mathrm{d} t^{2}+f(r)^{-1} \mathrm{~d} r^{2}+r^{2} \mathrm{~d} \Omega^{2}, \\
			\quad f(r)=1-\frac{2 M r^{2}}{\left(r^{2}+g^{2}\right)^{\frac{3}{2}}}+\frac{\alpha}{r} \ln \frac{r}{|\alpha|} ,\\
			\mathrm{~d} \Omega^{2}=d \theta^{2}+\sin ^{2} \theta d \phi^{2}.
		\end{array}
	\end{aligned}
\end{equation}

In 1965, Newman and Janis proposed Newman-Janis algorithm (NJA) to solve the rotation of regular black hole \cite{Newman:1965tw}, which has been widely used \cite{Kiselev:2002dx,Benavides-Gallego:2018odl,Kim:2019hfp,Liu:2020ola,Toshmatov:2017zpr,Shaikh:2019fpu,Xu:2020jpv,Kumar:2017qws,Xu:2016jod}. The metric of the rotating Bardeen black hole surrounded by perfect fluid dark matter \cite{Zhang:2020mxi} was obtained by \cite{Azreg-Ainou:2014pra,Azreg-Ainou:2014}

\begin{equation}\label{eq:ME}
	\begin{aligned}
		\begin{array}{c}
			\mathrm{d} s^{2}=-\left(1-\frac{2 \rho r}{\Sigma}\right) \mathrm{d} t^{2}+\frac{\Sigma}{\Delta_{r}} \mathrm{~d} r^{2}-\frac{4 a \rho r \sin ^{2} \theta}{\Sigma} \mathrm{d} t \mathrm{~d} \phi \\
			+\Sigma \mathrm{d} \theta^{2}+\sin ^{2} \theta\left(r^{2}+a^{2}+\frac{2 a^{2} \rho r \sin ^{2} \theta}{\Sigma}\right) \mathrm{d} \phi^{2},
		\end{array}
	\end{aligned}
\end{equation}
with
\begin{equation}
	\begin{aligned}
		\begin{array}{c}
			2 \rho=\frac{2 M r^{3}}{\left(r^{2}+g^{2}\right)^{\frac{3}{2}}}-\alpha \ln \frac{r}{|\alpha|}, \\
			\Sigma=r^{2}+a^{2} \cos ^{2} \theta, \\
			\Delta_{r}=r^{2}+a^{2}-\frac{2 M r^{4}}{\left(r^{2}+g^{2}\right)^{\frac{3}{2}}}+\alpha r \ln \frac{r}{|\alpha|}.
		\end{array}
	\end{aligned}
\end{equation}
For Eq.(\ref{eq:ME}), When $g$=0, it recovers to the rotating Bardeen black hole metric \cite{Bambi:2013ufa}, and when $g$=$\alpha$=0, it recovers to the Kerr black hole metric \cite{Kerr:1963ud}.

We know that the Bardeen black hole has no singularity and is regular everywhere, but the question is whether the perfect fluid dark
matter outside the black hole will affect this characteristic. Therefore, the curvature scalar of the black hole was calculated as \cite{Zhang:2020mxi}
\begin{equation}
	\begin{array}{c}
		R=\frac{6 M g^{2}\left(4 g^{2}-r^{2}\right)}{\left(g^{2}+r^{2}\right)^{\frac{7}{2}}}-\frac{\alpha}{r^{3}} ,
	\end{array}
\end{equation}
\begin{equation}
	\begin{array}{c}
		R_{\mu \nu} R^{\mu \nu}=\frac{18 M^{2} g^{4}\left(8 g^{4}-4 g^{2} r^{2}+13 r^{4}\right)}{\left(g^{2}+r^{2}\right)^{7}}+\frac{5 \alpha^{2}}{2 r^{6}}\\-\frac{6 M g^{2}\left(2 g^{2}+7 r^{2}\right) \alpha}{r^{3}\left(g^{2}+r^{2}\right)^{\frac{7}{2}}} ,
	\end{array}
\end{equation}

when $r$=0, the results obtained in the above equation do not diverge, indicating that the Bardeen black hole surrounded by perfect fluid dark matter has no singularity.
\subsection{Horizons}

The metric (\ref{eq:ME}) is singular at $\Delta$= 0 and corresponds to the horizon of black hole
\begin{equation}\label{eq:HO}
	\begin{aligned}
		\begin{array}{c}
			\Delta_{r}=r^{2}+a^{2}-\frac{2 M r^{4}}{\left(r^{2}+g^{2}\right)^{\frac{3}{2}}}+\alpha r \ln \frac{r}{|\alpha|}=0.
		\end{array}
	\end{aligned}
\end{equation}
For Eq.(\ref{eq:HO}), we can obtain the Cauchy horizon $r_H^-$ and event horizon $ r_H^+ $ of the rotating Bardeen black hole surrounded by perfect fluid dark matter. The behavior of the event horizon with different values of rotation parameter $a$, magnetic charge parameter $g$ and dark matter parameter $\alpha$ is shown in Figs.~\ref{fig:horizon1}, \ref{fig:horizon2}.
When the two event horizons of a black hole coincide, the black hole is an extreme black hole with a specific critical rotation parameter a=$a_E$, and the extreme black hole satisfies
\begin{equation} \label{eq:DH}
	\begin{array}{c}
		\alpha +\alpha  \ln \left(\frac{r}{\left| \alpha \right| }\right)+\frac{6 M r^5}{\left(g^2+r^2\right)^{5/2}}-\frac{8 M r^3}{\left(g^2+r^2\right)^{3/2}}+2 r=0.
	\end{array}
\end{equation}

From Eq.(\ref{eq:HO}) and (\ref{eq:DH}), we can obtain the critical value of $a_E$ for given $g$ and $\alpha$ by numerical method, and then we draw a three-dimensional contour plane with $\Delta$=0 in Fig.~\ref{fig:horizon3}. By analyzing Fig.~\ref{fig:horizon3}, we can find the following conclusions.

\begin{itemize} 	
	\item{In order to get an extreme black hole, there's a range of values for $g$ and $\alpha$, and it's only within that range that you find $a_E$.}
	\item {The three planes $g$=0, $\alpha$=0, $a$=0 and the contour plane of $\Delta$=0 form a spatial region.}
		\begin{enumerate}
			\item[1)] The vertical coordinate $a$ of the coordinate point in this spatial area satisfies $a <a_E$, and $\Delta$=0 has two roots representing non-extreme black holes with two different horizons.
			\item[2)] The vertical coordinate $a$ of coordinate points outside this spatial area satisfies $a>a_E$, $\Delta$=0 has no root. There is no horizon in this particular case.
			\item[3)] When the vertical coordinate $a$ of the coordinate point on the contour plane determined by $\Delta$=0 satisfies $a=a_E$, and there is a root, the black hole is an extreme black hole.
		\end{enumerate}
	\end{itemize}

\begin{figure*}
	\begin{tabular}{c c}
		\includegraphics[scale=0.58]{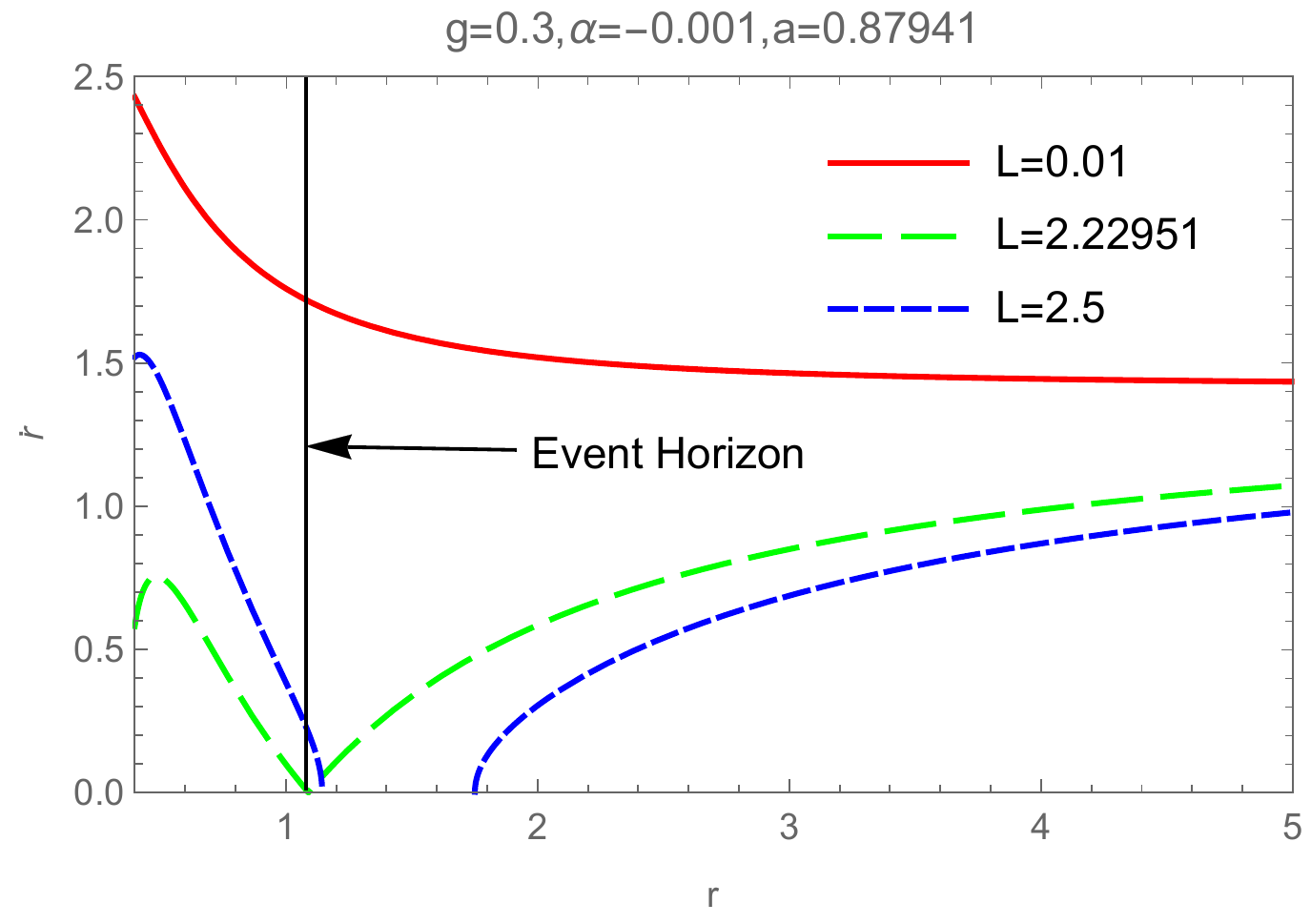}\hspace{-0.2cm}
		&\includegraphics[scale=0.58]{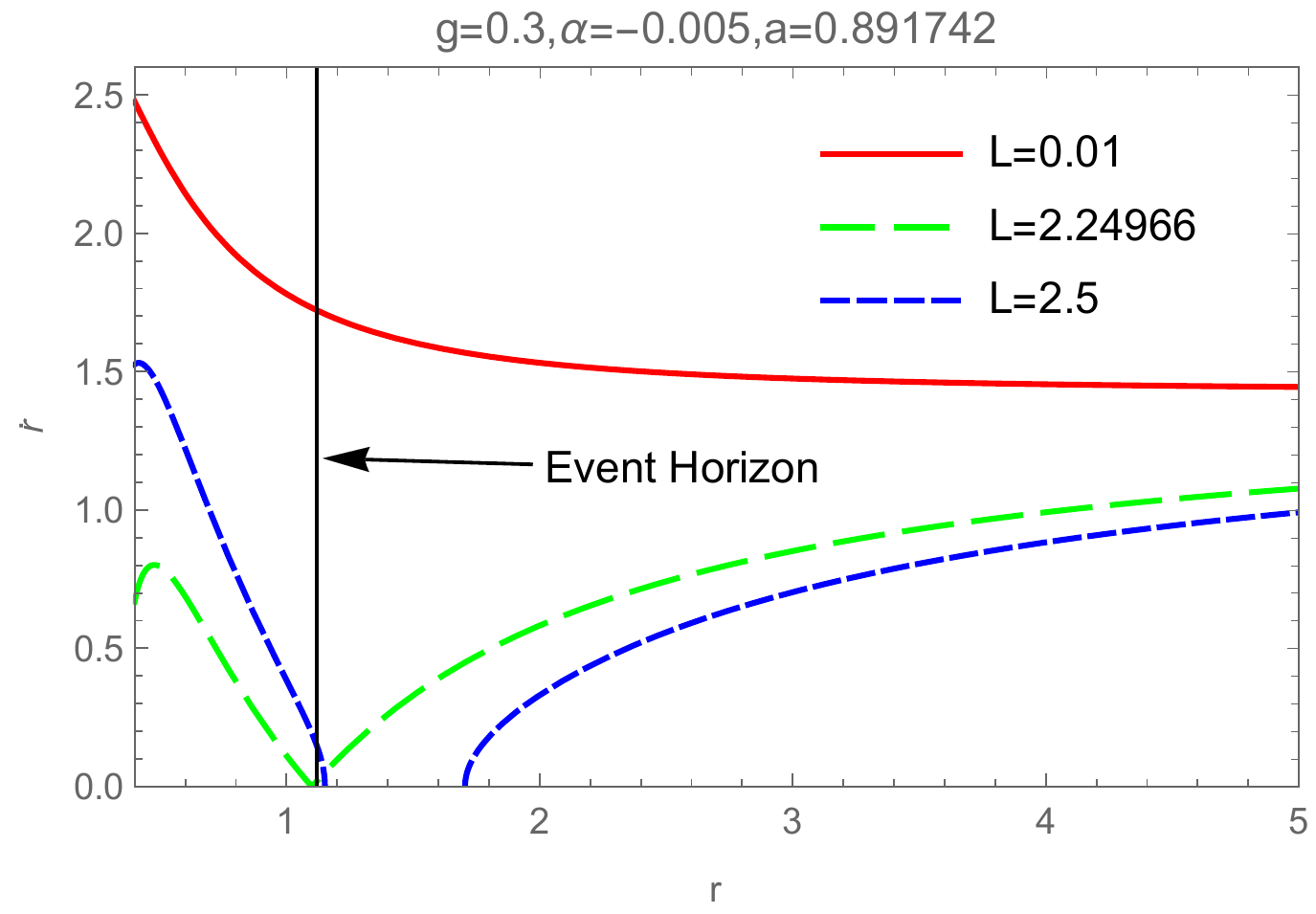}\\
		\includegraphics[scale=0.58]{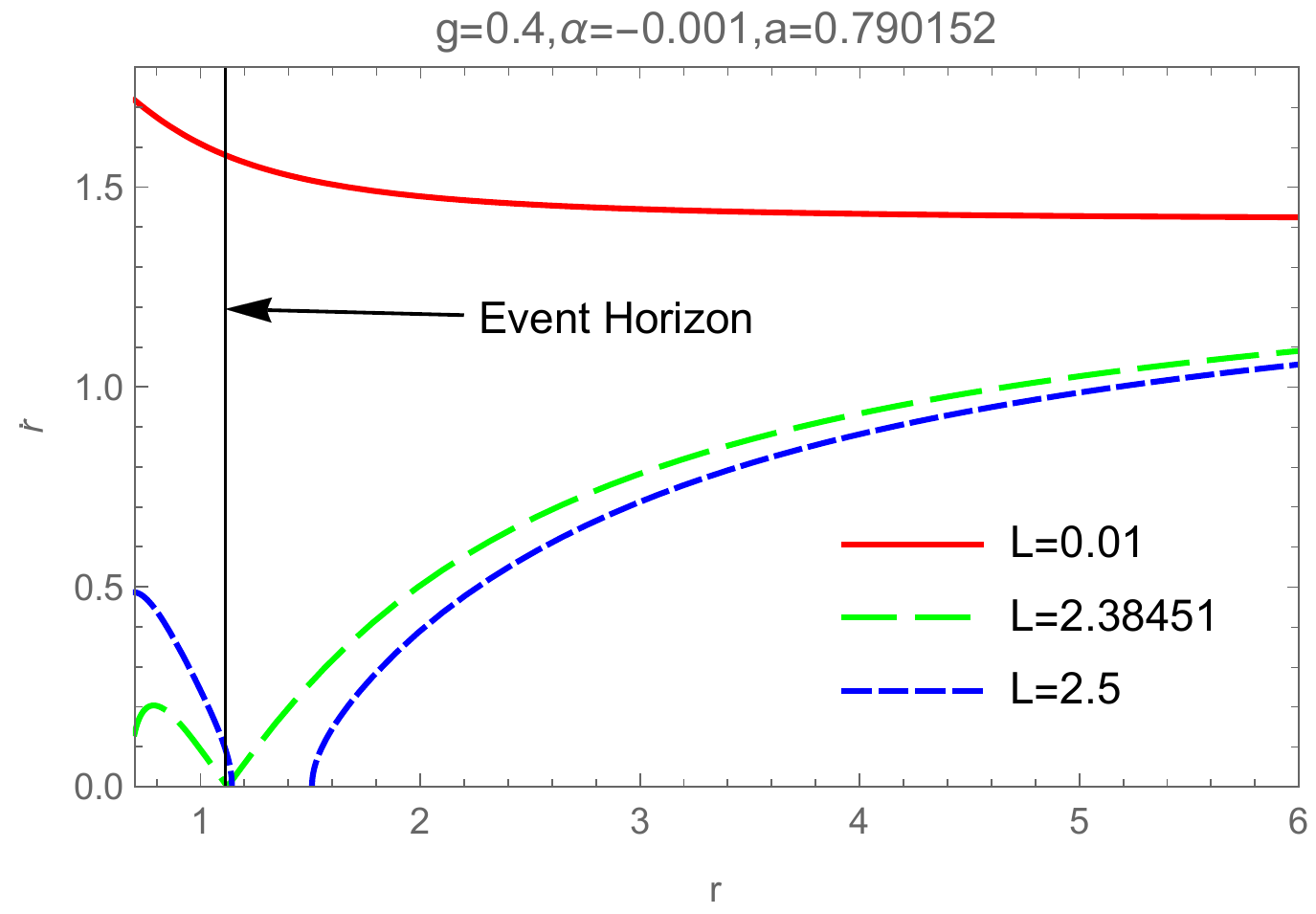}\hspace{-0.2cm}
		&\includegraphics[scale=0.58]{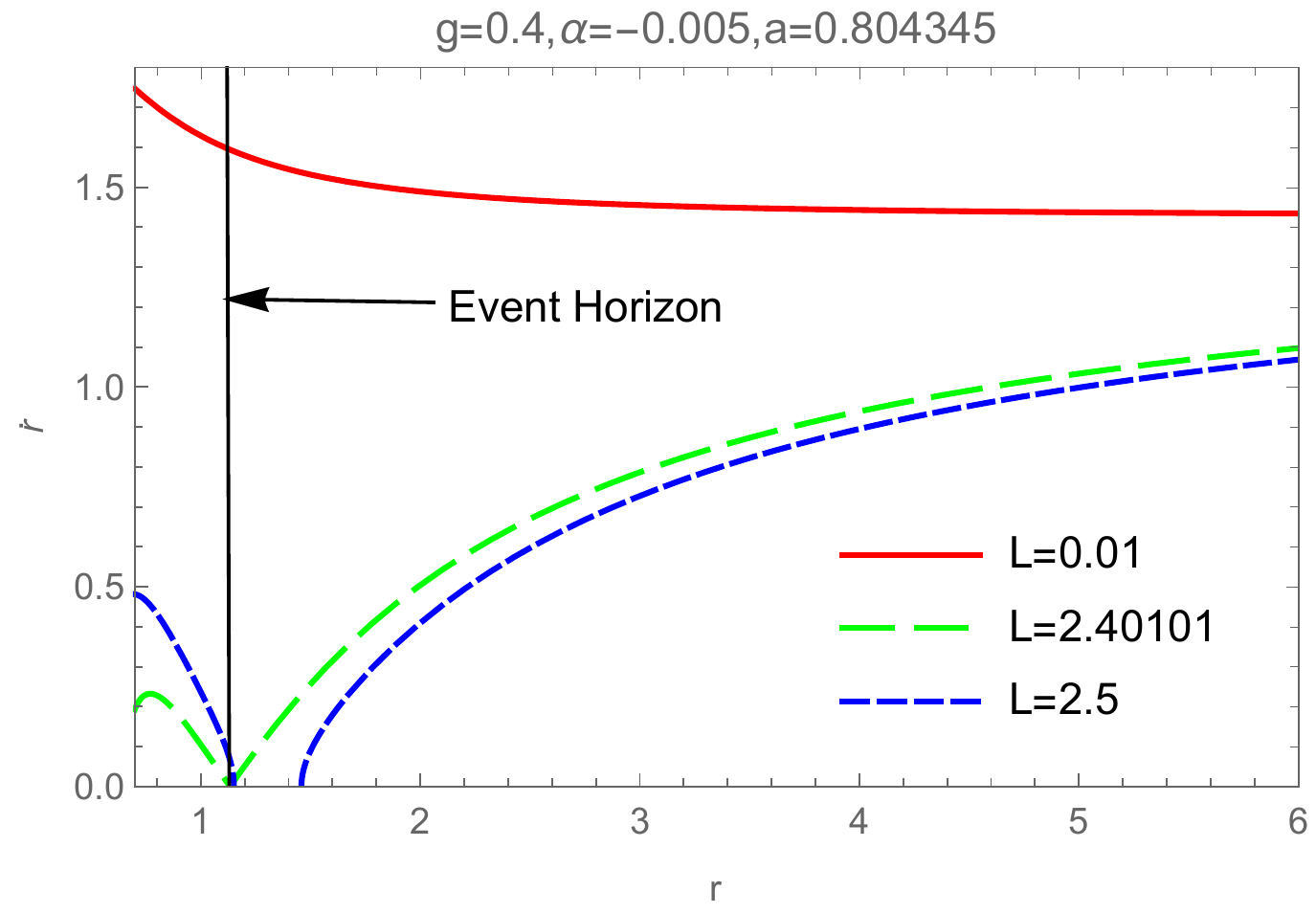}
	\end{tabular}
	\caption{The behavior of $\dot{r}$  vs $r$ for extremal black hole.}\label{fig:dotr}
\end{figure*}

\begin{figure*}
	\begin{tabular}{c c}
		\includegraphics[scale=0.64]{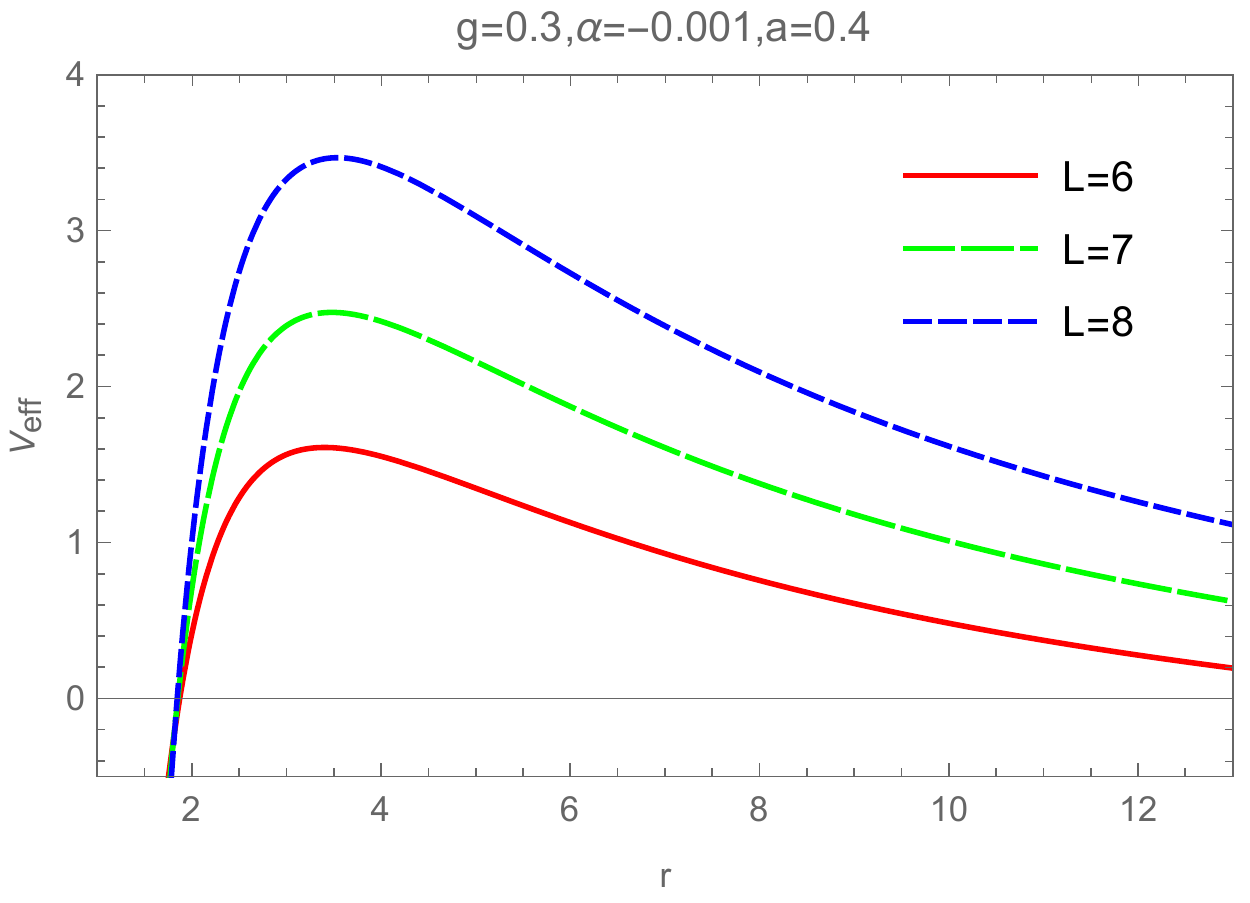}\hspace{-0.2cm}
		&\includegraphics[scale=0.64]{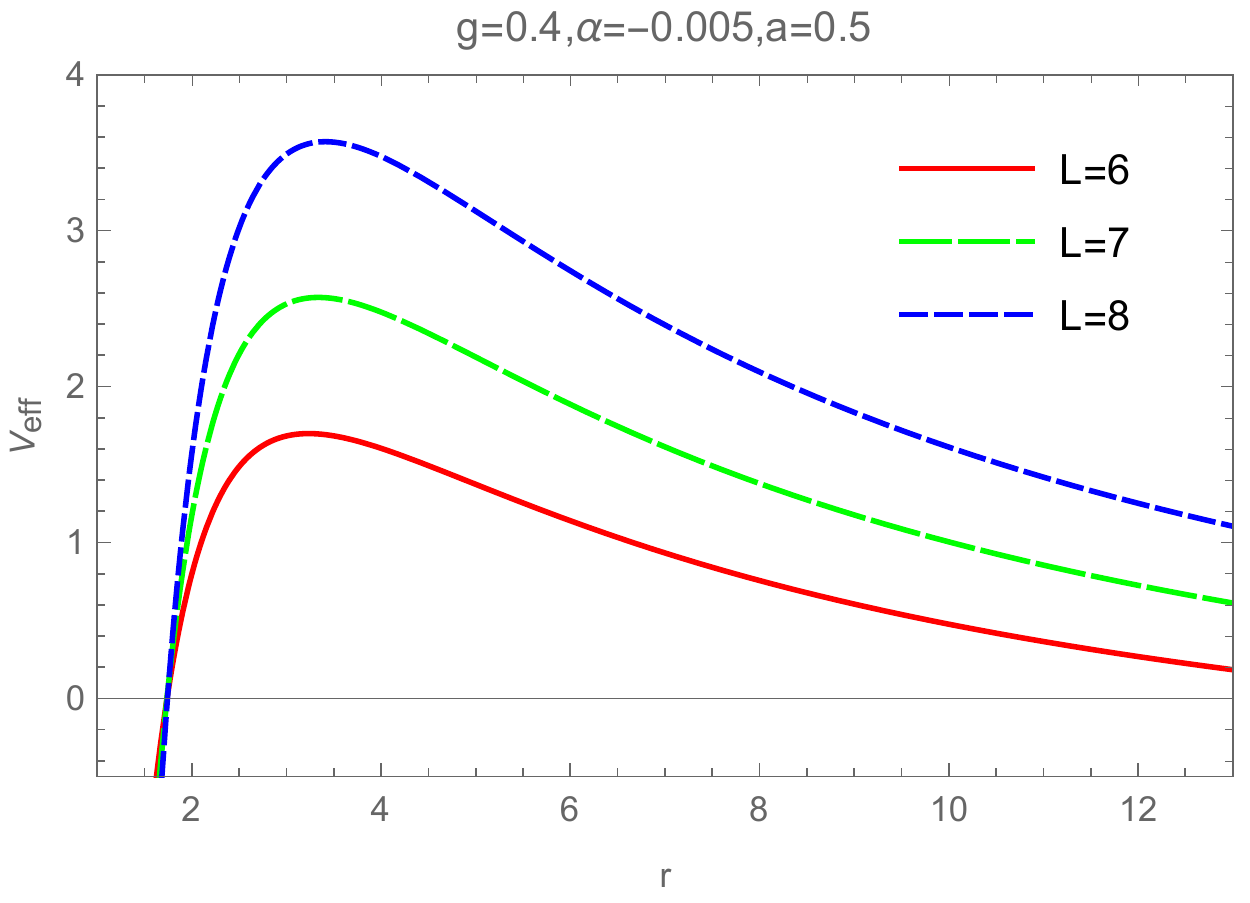}\\
		\includegraphics[scale=0.64]{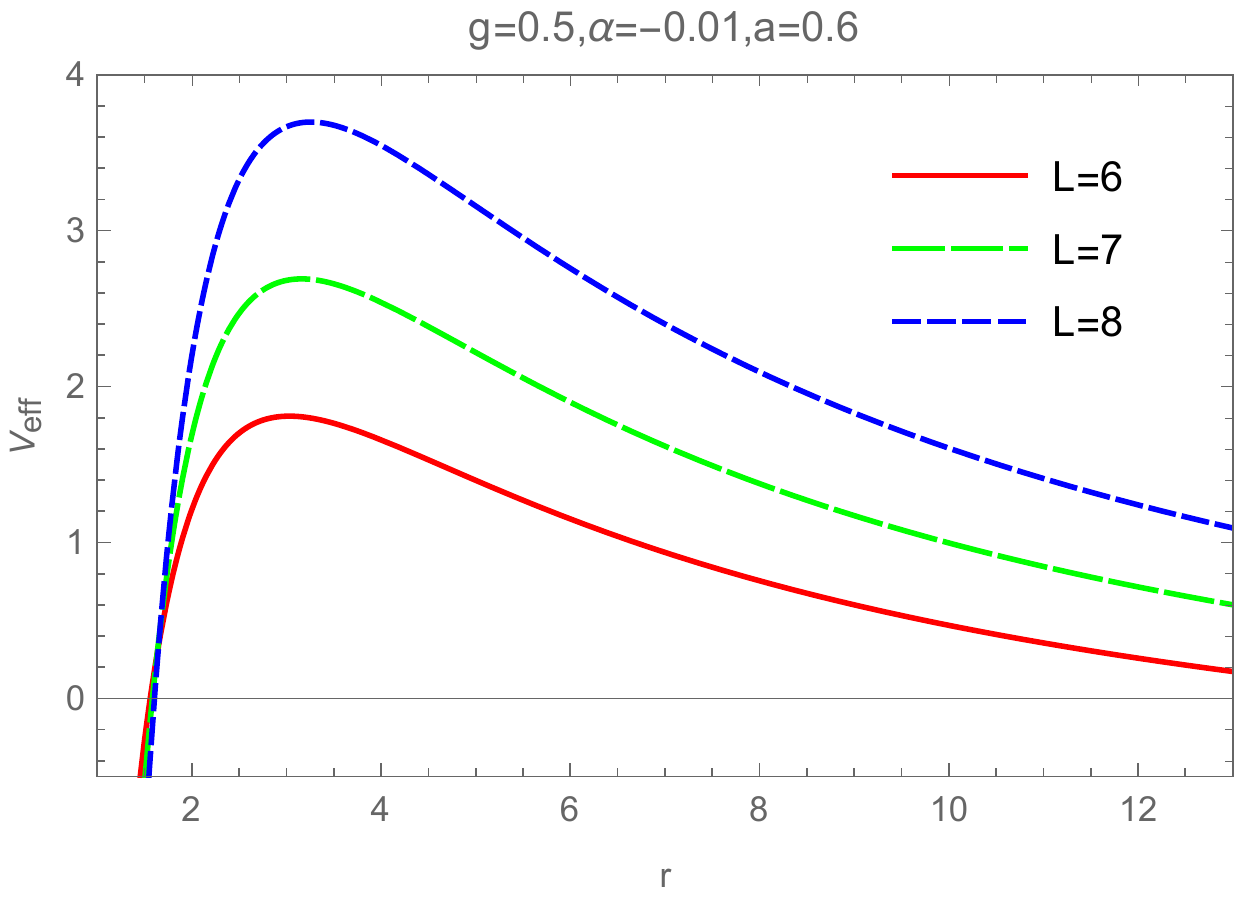}\hspace{-0.2cm}
		&\includegraphics[scale=0.64]{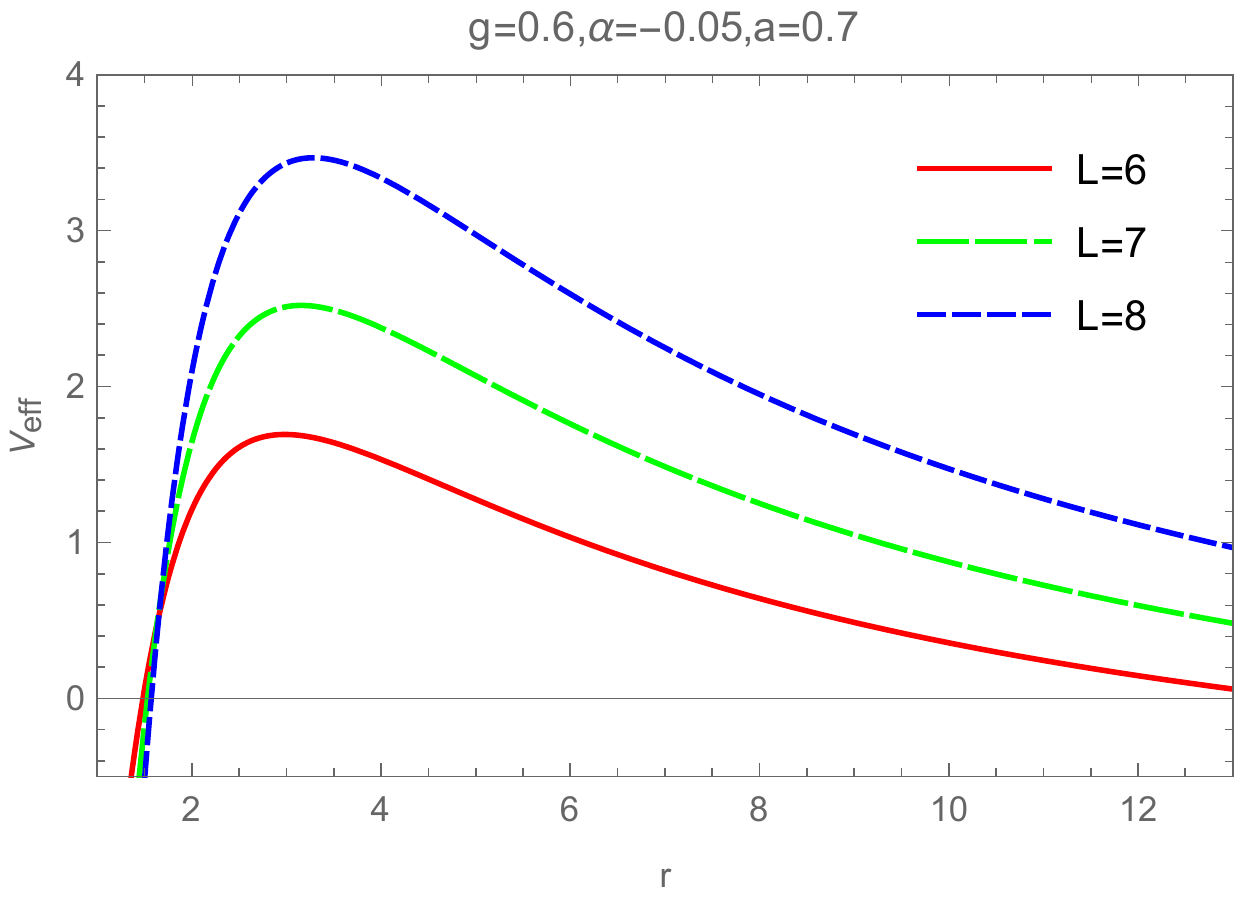}
	\end{tabular}
	\caption{Plot showing the behavior of $V_{eff}$ vs $r$ for different angular momentum $L$.}\label{fig:Veff}
\end{figure*}

\begin{table}
	\begin{center}
		\caption{Limit values of angular momentum for different extreme values of a rotating Bardeen black hole surrounded by perfect fluid dark matter.}\label{table1}
		\begin{tabular}{| c c | c | c | c | c | c |}
			\hline
			& $g$	& $\alpha$  & $a_{E}$    & $r^{E}_{H}$  & $L_{2}$    & $L_{1}$ \\
			\hline
			& 0 	& 0         & 1.0        & 1.00000  	&-4.82843    & 2.00000  \\
			& 0.2 	& -0.001    & 0.946305   & 1.05138  	&-4.80050    & 2.11443  \\
			& 0.2 	& -0.005    & 0.957301   & 1.06245 		&-4.86331    & 2.13645  \\
			& 0.2  	& -0.01     & 0.968335   & 1.07406		&-4.92993    & 2.15967  \\
			& 0.3 	& -0.001    & 0.879410   & 1.08963 		&-4.74191    & 2.22951  \\
			& 0.3 	& -0.005    & 0.891742   & 1.10041   	&-4.80574    & 2.24966  \\
			& 0.3 	& -0.01     & 0.904100   & 1.11174      &-4.87334    & 2.27117  \\
			& 0.4 	& -0.001    & 0.790152   & 1.12240  	&-4.66015    & 2.38451  \\
			& 0.4 	& -0.005    &0.804345    & 1.13326 		&-4.72558    & 2.40101  \\
			& 0.4  	& -0.01     & 0.818529   & 1.14465		&-4.79469    & 2.41923  \\
			
			\hline
		\end{tabular}
	\end{center}
\end{table}
\begin{table}
	\begin{center}
		\caption{The limiting values of angular momentum for different non-extremal cases of rotating Bardeen black hole surrounded by perfect fluid dark matter.}\label{table2}
		\begin{tabular}{| c c | c | c | c | c | c | c |}
			\hline
			& $g$   & $\alpha$& $a$  & $r^{+}_{H}$  & $r^{-}_{H}$   & $L_{4}$    & $L_{3}$ \\
			\hline
			& 0 	& 0        & 7     & 1.71414  	& 0.285857		&-4.60768    & 3.09545 \\
			& 0.2 	& -0.001   & 0.5   & 1.83631  	& 0.295480   	&-4.45677    & 3.40098  \\
			& 0.2	& -0.005   & 0.6   & 1.79167  	& 0.360599		&-4.59112    & 3.29584  \\
			& 0.2 	& -0.01    & 0.7   & 1.72855  	& 0.442073		&-4.72735    & 3.17179  \\
			& 0.3 	& -0.001   & 0.5   & 1.78571  	& 0.398506		&-4.44190    & 3.36117  \\
			& 0.3  	& -0.005   & 0.6   & 1.73538  	& 0.475478		&-4.57957    & 3.24747  \\
			& 0.3   & -0.01    & 0.7   & 1.66256  	& 0.571569		&-4.71670    & 3.10970  \\
			& 0.4 	& -0.001   & 0.5   & 1.70684  	& 0.519312   	&-4.42619    & 3.29900  \\
			& 0.4	& -0.005   & 0.6   & 1.64502  	& 0.613882		&-4.56309    & 3.16911  \\
			& 0.4 	& -0.01    & 0.7   & 1.54990  	& 0.737884		&-4.70153    & 3.00160  \\
			
			\hline
		\end{tabular}
	\end{center}
\end{table}

\begin{figure*}
	\begin{tabular}{c c c c}
		\includegraphics[scale=0.5]{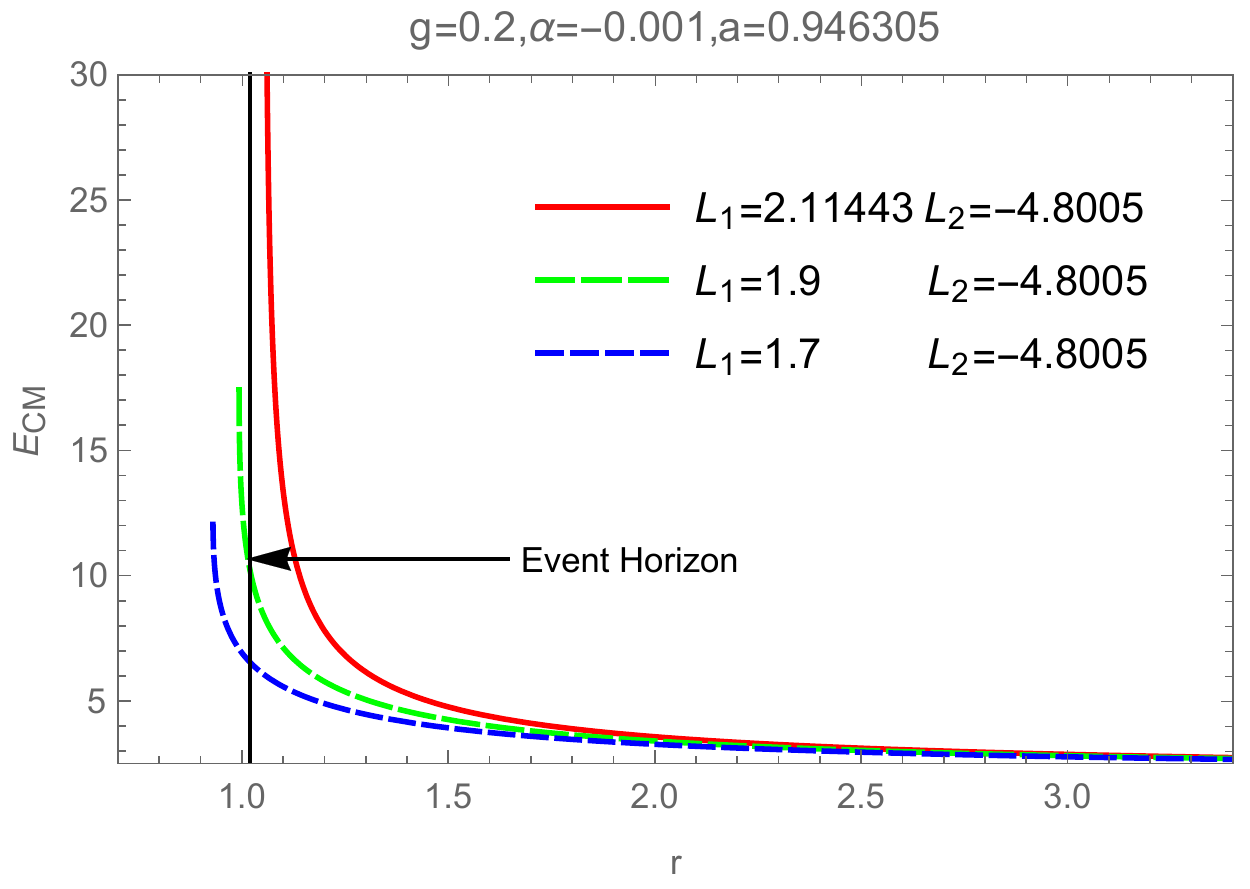}\hspace{-0.2cm}
		\includegraphics[scale=0.5]{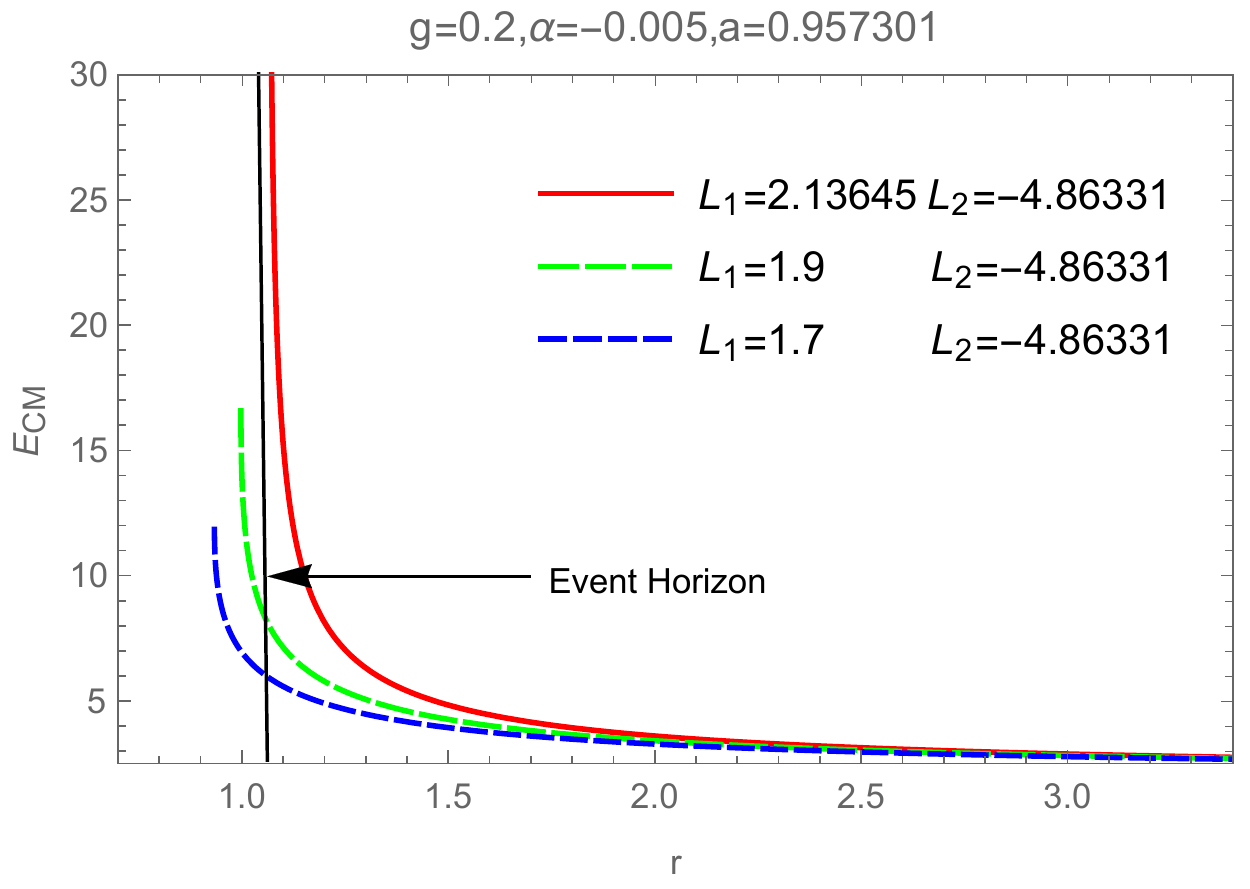}\hspace{-0.2cm}
		&\includegraphics[scale=0.5]{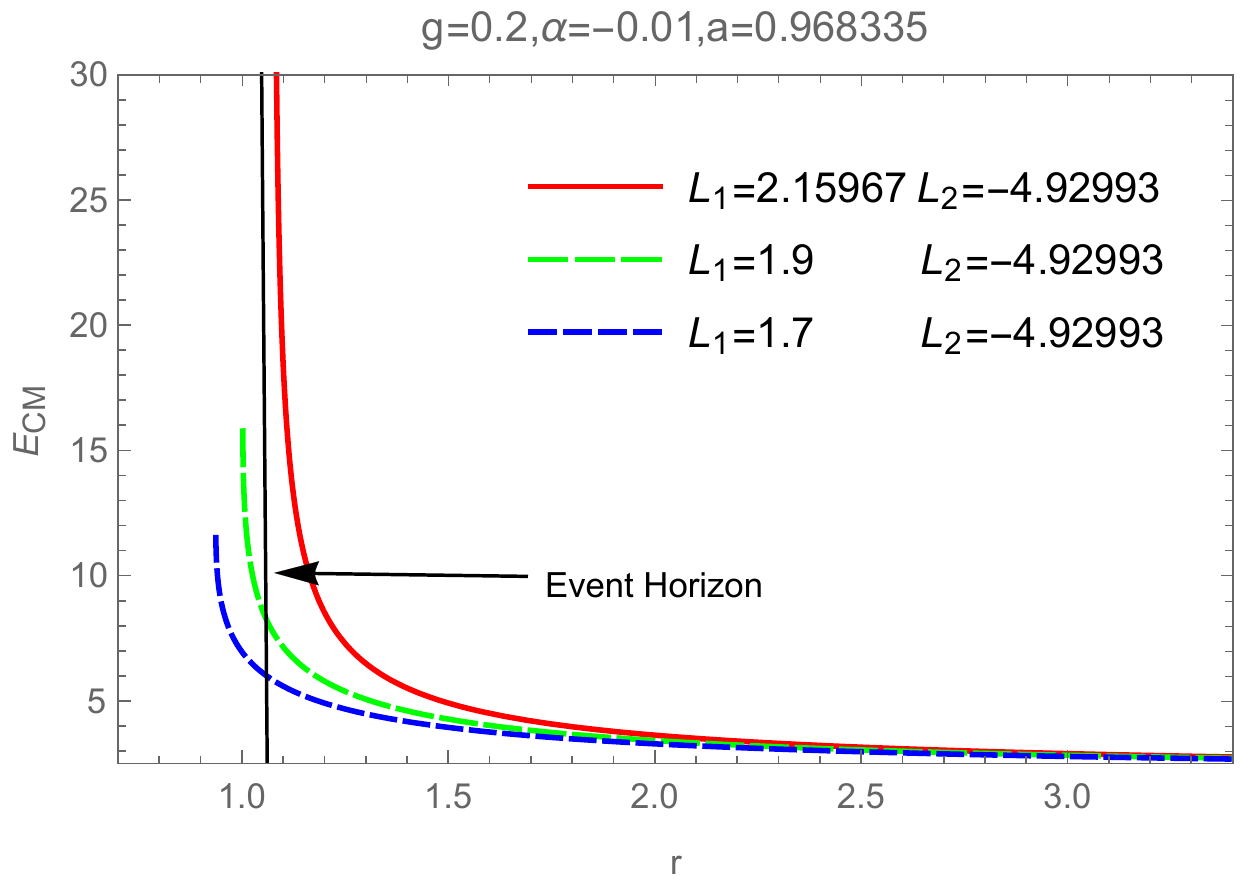}\\
		\includegraphics[scale=0.5]{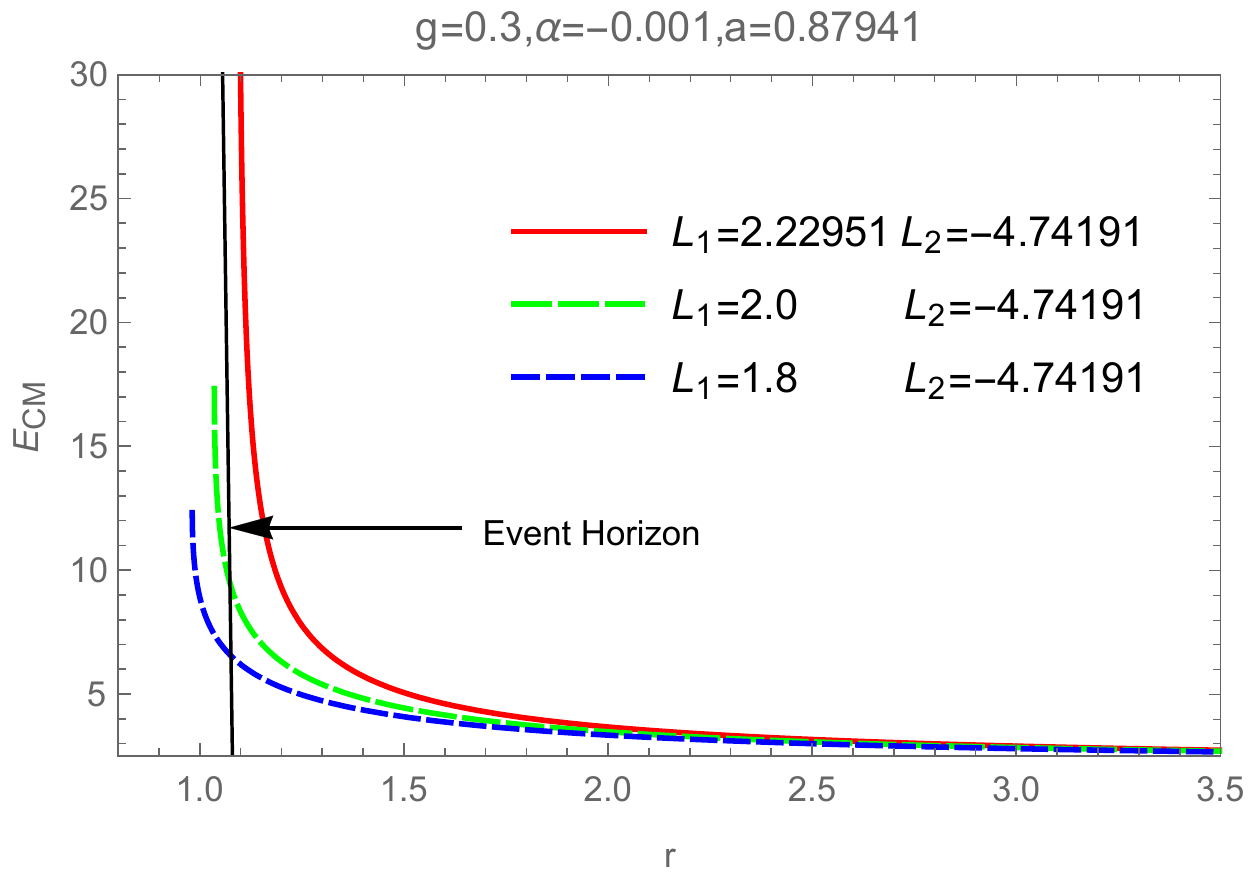}\hspace{-0.2cm}
		\includegraphics[scale=0.5]{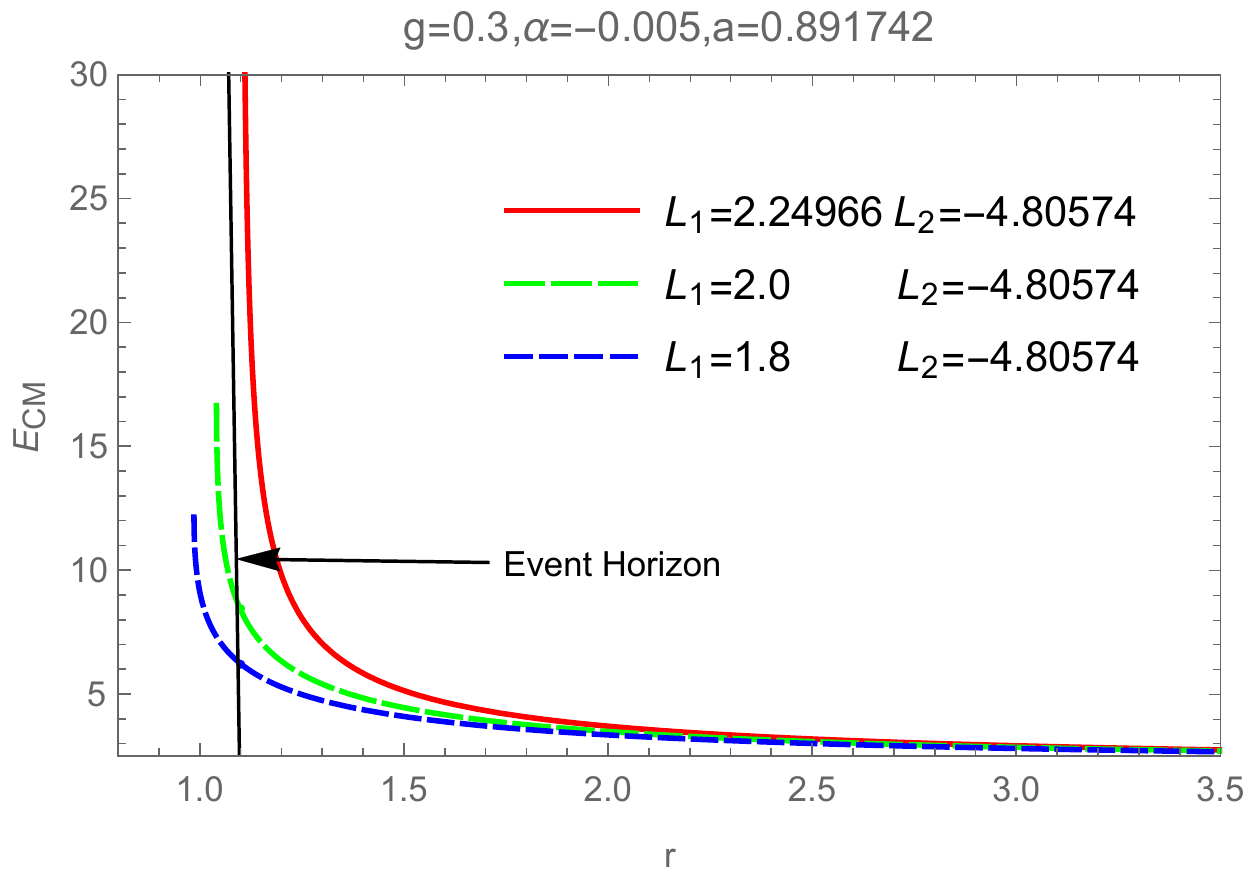}\hspace{-0.2cm}
		&\includegraphics[scale=0.5]{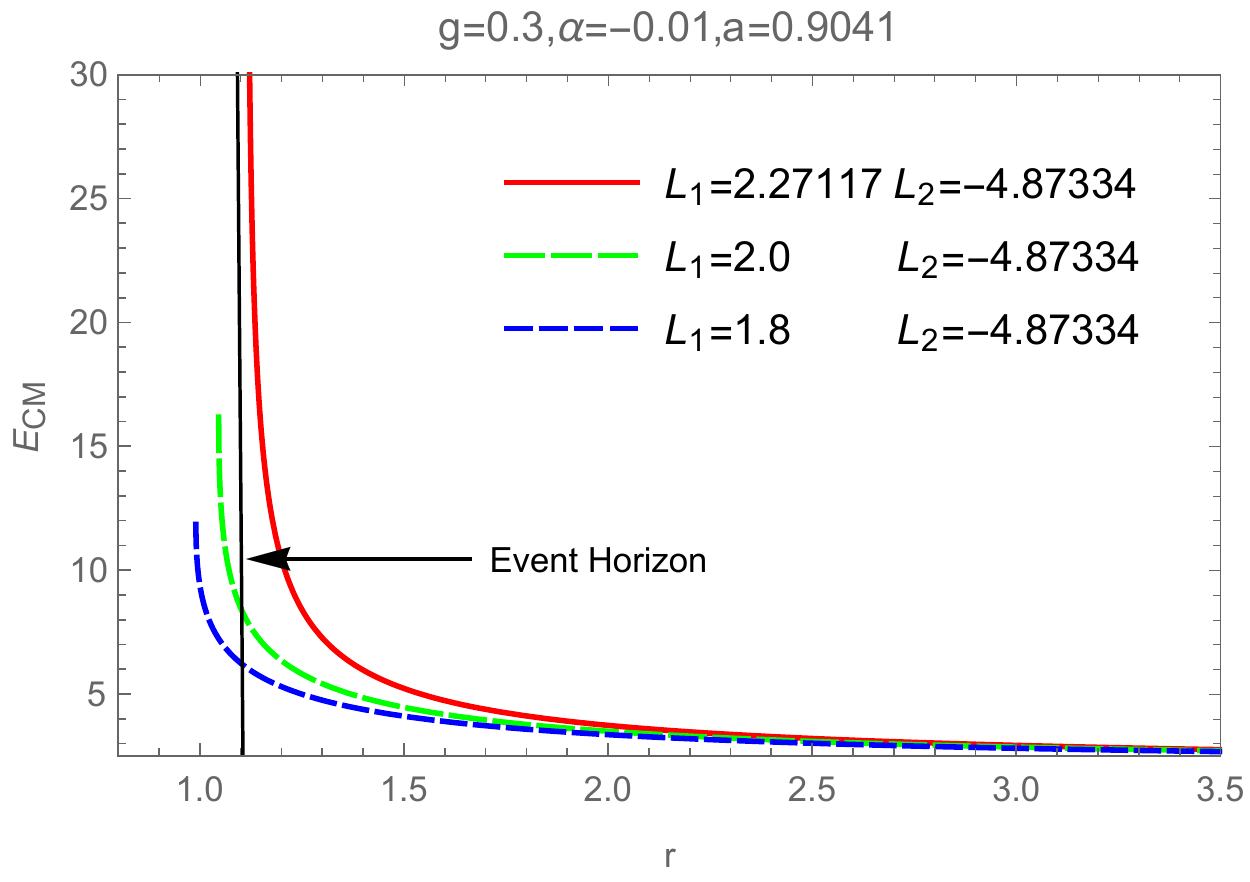}\\
		\includegraphics[scale=0.5]{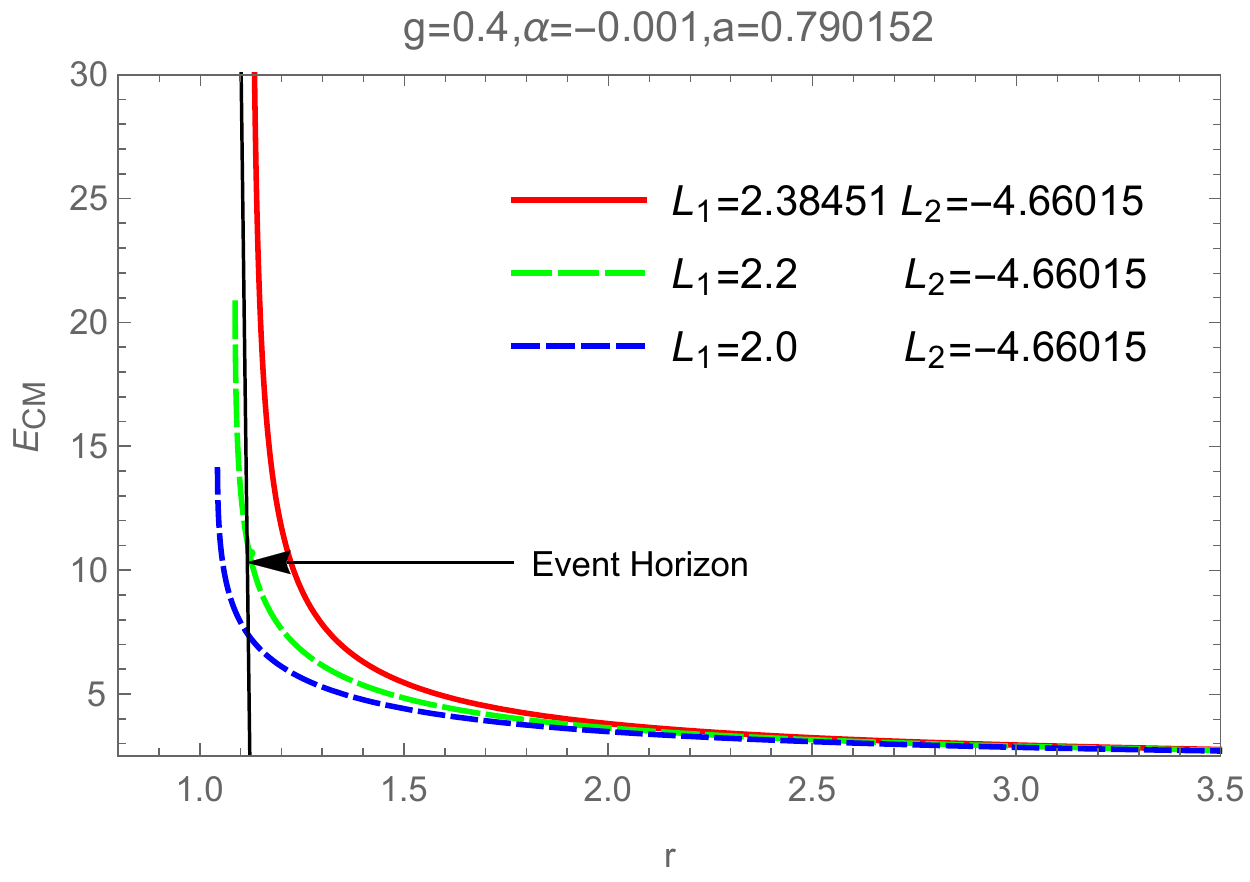}\hspace{-0.2cm}
		\includegraphics[scale=0.5]{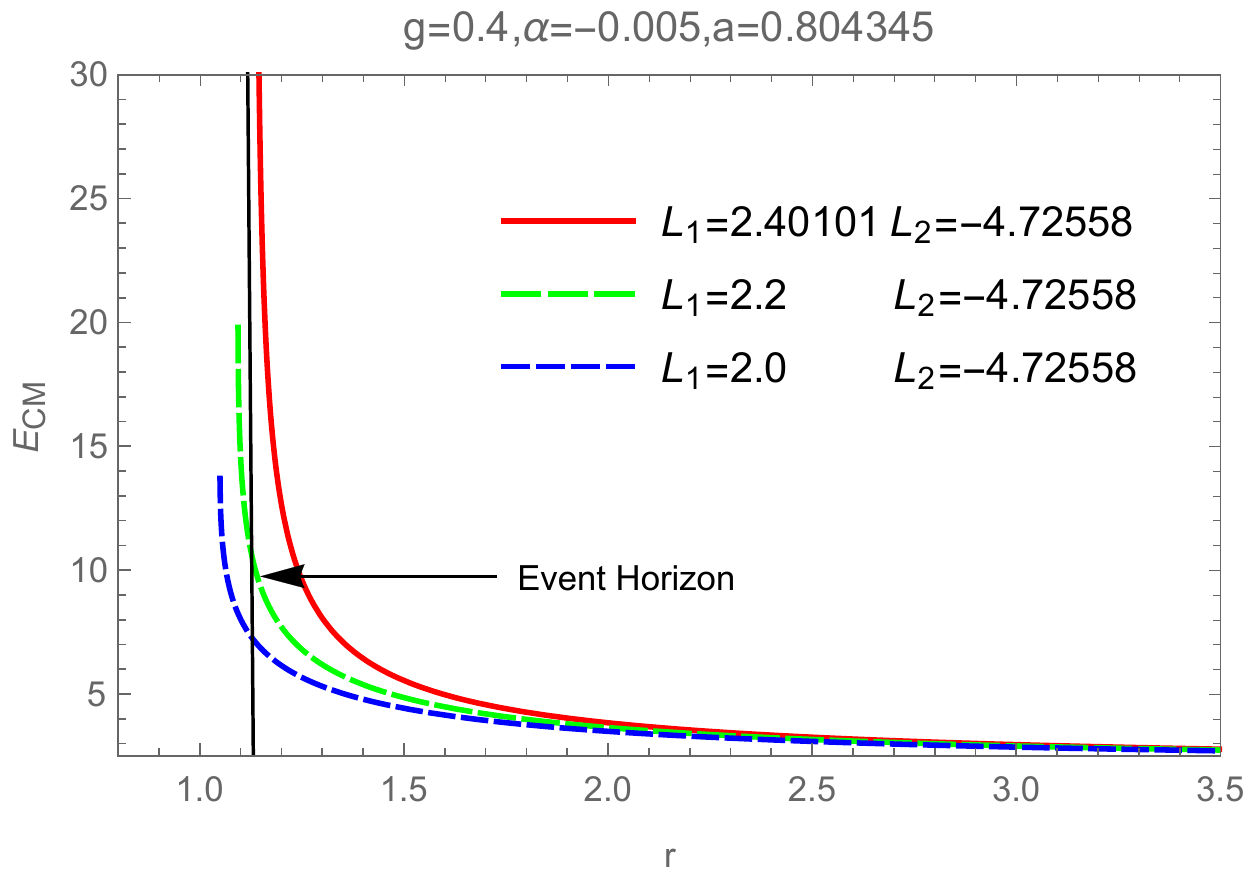}\hspace{-0.2cm}
		&\includegraphics[scale=0.5]{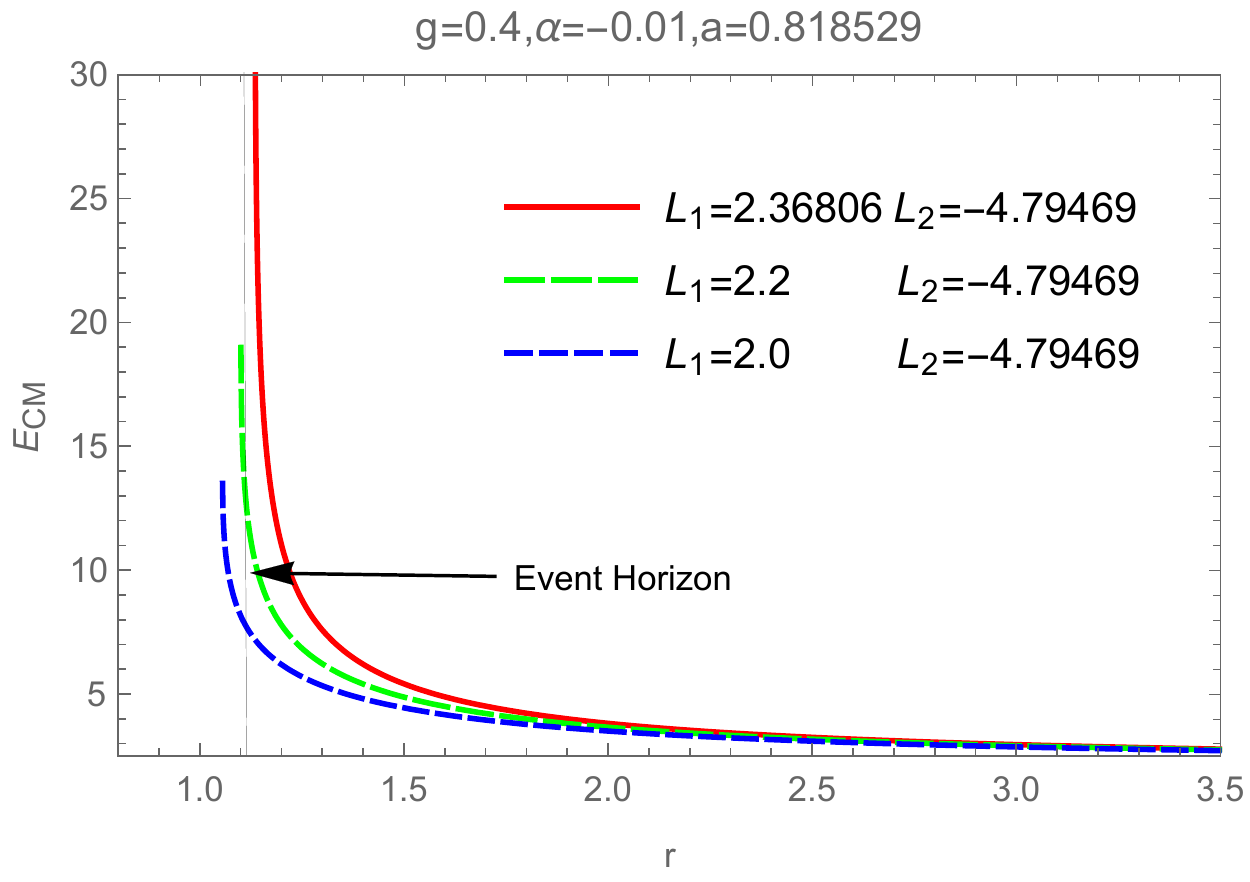}
	\end{tabular}
	\caption{Plot showing the behavior of $E_{CM}$ vs $r$ for extremal black hole.}\label{fig:Ecm1}
\end{figure*}

\begin{figure*}
	\begin{tabular}{c c c c}
		\includegraphics[scale=0.455]{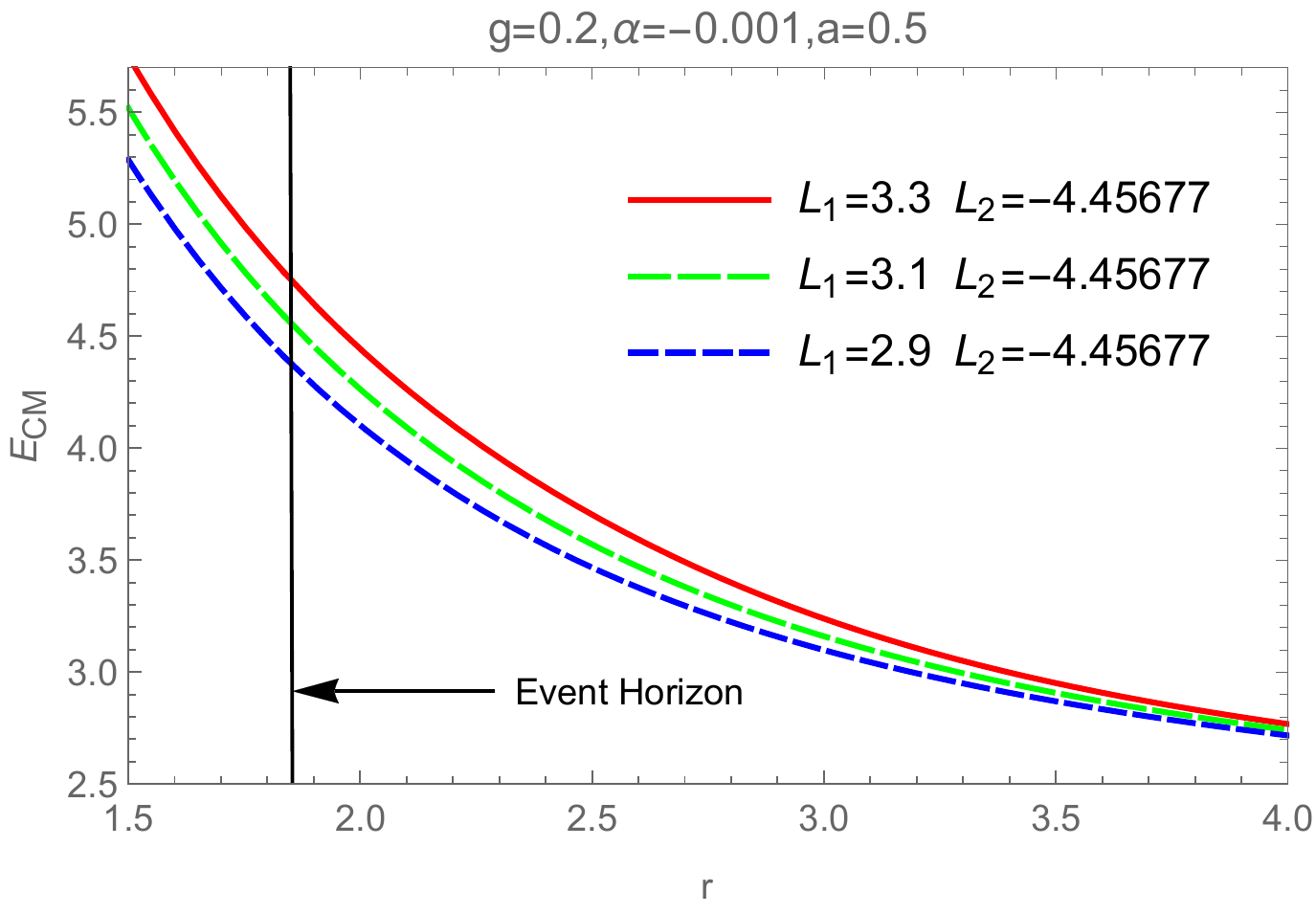}\hspace{-0.2cm}
		\includegraphics[scale=0.455]{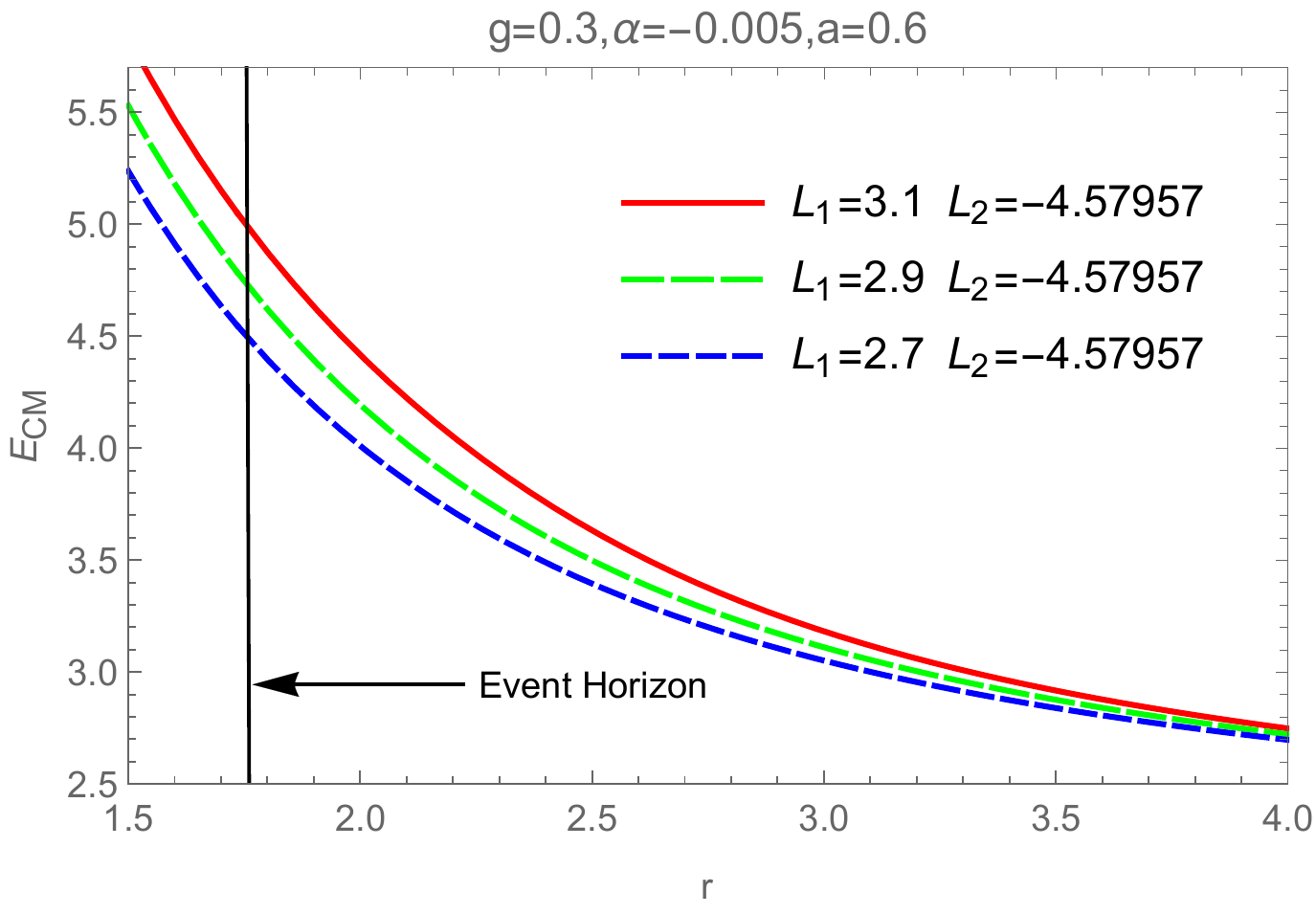}\hspace{-0.2cm}
		&\includegraphics[scale=0.455]{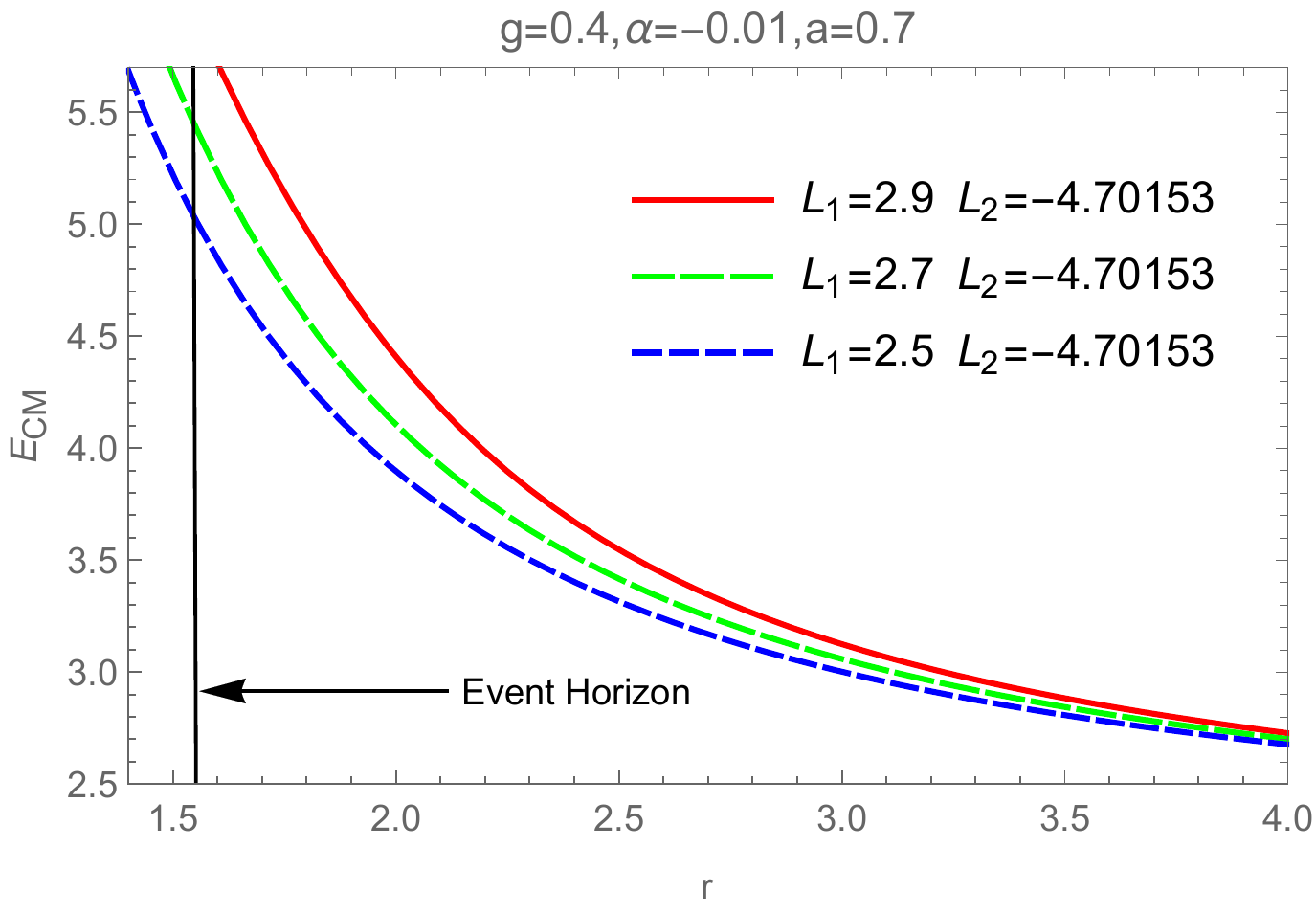}
	\end{tabular}
	\caption{Plot showing the behavior of $E_{CM}$ vs $r$ for non-extremal black hole.}\label{fig:Ecm2}
\end{figure*}

\section{EQUATIONS OF MOTION AND THE EFFECTIVE POTENTIAL}
\label{sec:3}
In this section, we study the equations of motion for a time-like particle with rest mass $m_0$ in the background of a rotating Bardeen black hole surrounded by perfect fluid dark matter. We consider the particle moving on the equatorial plane of the black hole ($\theta =\pi/2$, $ \dot{\theta } =0$), and the generalized momentum of the particle in the space-time of the rotating Bardeen black hole surrounded by perfect fluid dark matter is expressed as follows

\begin{equation}\label{eq:PT}
	\begin{array}{c}
		\begin{aligned}
			P_{t} &=g_{t t} \dot{t}+g_{t \phi} \dot{\phi} ,
		\end{aligned}
	\end{array}
\end{equation}
\begin{equation}\label{eq:PF}
	\begin{array}{c}
		\begin{aligned}
			P_{\phi} &=g_{\phi \phi} \dot{\phi}+g_{t \phi} \dot{t},
		\end{aligned}
	\end{array}
\end{equation}

where $P_t$ and $P_\phi $ are motion constants corresponding to the particle energy $E$ per unit mass and the angular momentum $L$ of the particle per unit mass parallel to the spin axis of the black hole, respectively. So, we have $P_t=-E$, $P_\phi =L$. By simplifying the conditions for Eq.(\ref{eq:PT}), (\ref{eq:PF}) and $P_{\nu} P^{\nu}=-m_{0}^{2}$, the four-velocity of the particle can be obtained to have the following form:

\begin{equation}\label{eq:T}
	\begin{array}{c}
		\begin{aligned}
			\dot{t} &=\frac{T}{\Delta_r}E-\frac{2a\rho }{r\Delta_r}L,
		\end{aligned}
	\end{array}
\end{equation}

\begin{equation}\label{eq:F}
	\begin{array}{c}
		\begin{aligned}
			\dot{\phi} &=\frac{2a\rho }{r\Delta_r}E+\frac{(1-\frac{2\rho}{r}) }{\Delta_r}L,
		\end{aligned}
	\end{array}
\end{equation}
\begin{equation}\label{eq:R}
	\begin{array}{c}
		\begin{aligned}
			\dot{r} &=\pm\frac{\sqrt{TE^2-(1-\frac{2\rho}{r})L^2-(\frac{4a\rho}{r})EL-\Delta _rm_0^2 } }{r} ,
		\end{aligned}
	\end{array}
\end{equation}
where $T=r^2+a^2+\frac{2a^2\rho}{r} $. In Eq.(\ref{eq:R}), the sign $+$ indicates the outgoing geodesic, and the sign $-$ corresponds to the incoming geodesic. To obtain the angular momentum range of the falling particle, we need to calculate the effective potential. Consider the relationship between the radial motion described by Eq.(\ref{eq:R}) and the effective potential
\begin{equation}
	\begin{array}{c}
		\begin{aligned}
			\frac{1}{2} \dot{r}^{2}+V_{{eff }}=0,
		\end{aligned}
	\end{array}
\end{equation}
so we have an effective potential expression
\begin{equation}
	\begin{array}{c}
		\begin{aligned}
			V_{eff} =-\frac{TE^2-(1-\frac{2\rho }{r} )L^2-\frac{4a\rho }{r}EL-\Delta _r }{2r^2}.
		\end{aligned}
	\end{array}
\end{equation}

The maximum and minimum angular momentum of the falling particle satisfy the following equation
\begin{equation}\label{eq:L}
	\begin{array}{c}
		\begin{aligned}
			V_{e f f}=0 \quad \text { and } \quad \frac{d V_{e f f}}{d r}=0.
		\end{aligned}
	\end{array}
\end{equation}

From the constraints of the above equation, we obtain the limiting values $L_{min}$ and $L_{max}$ of the angular momentum of the incident particles for extreme (table \ref{table1}) and non-extreme (table \ref{table2}) black holes. There is  $dt/d\tau >0$ for equation (\ref{eq:T}) because the geodesic of the particle is time-like
\begin{equation}
	\begin{array}{c}
		\begin{aligned}
			(r^2+a^2+\frac{2a^2\rho }{r})E-\frac{2a\rho }{r} L\ge 0,
		\end{aligned}
	\end{array}
\end{equation}
the above equation can be obtained when the condition $r\to r_{H}^{E}$ is satisfied
\begin{equation}\label{eq:L1}
	\begin{array}{c}
		\begin{aligned}
			E-\Omega_{H} L \geq 0,
		\end{aligned}
	\end{array}
\end{equation}
where $\Omega_{H}$ is the angular velocity of the event horizon of the black hole, which is given by
\begin{equation}
	\begin{array}{c}
		\begin{aligned}
			\Omega_{H}=\frac{a}{r_{H}^{E}+a^{2}}.
		\end{aligned}
	\end{array}
\end{equation}
The critical angular momentum $L_c =E/\Omega_{H}$ is obtained when we take the equal sign of Eq.(\ref{eq:L1}). In Fig.~\ref{fig:dotr} we give the curves of $\dot{r} $ vs $r$ for different values of $L$, $a$ and $g$, $\alpha$. As can be seen from the figure, if the particle has a large angular momentum $L>L_c$, geodesics don't fall into black holes. On the other hand, if the angular momentum $L<L_c$ the geodesic always falls in the black hole, and when $L = L_c$, the geodesic falls right at the black hole horizon. The effective potential ($V_{eff}$) varies with radius ($r$) as shown in Fig.~\ref{fig:Veff}.

\section{NEAR HORIZON COLLISION IN ROTATING BARDEEN BLACK HOLE IN PERFECT FLUID DARK MATTER}
\label{sec:4}
In this part, we calculate the CM energy of two time-like particles outside the black hole, and discuss the conditions under which the CM energy of two particles colliding at the horizon of the black hole will diverge.

For two uncharged time-like particles falling from rest at infinity of the equatorial plane of the black hole, with the mass $m_1=m_2=m_0$, energy $E_1=E_2=1$, angular momentum $L_1$ and $L_2$, respectively. The CM energy of the two particles at the radial coordinate $r$ of the equatorial plane is expressed as
\begin{equation}
	\begin{array}{c}
		\begin{aligned}
			E_{C M}=\sqrt{2}m_{0}  \sqrt{1-g_{\mu \nu} u_{(1)}^{\mu} u_{(2)}^{\nu}} ,
		\end{aligned}
	\end{array}
\end{equation}
where $u_{(1)}^{\mu}$ and $u_{(2)}^{\nu}$ are the four-velocities of two particles, respectively. Therefore, we substitute equations (\ref{eq:T}), (\ref{eq:F}) and (\ref{eq:R}) into the above equation to simplify

\begin{equation}\label{eq:Ecm}
	\begin{array}{c}
		\begin{aligned}
			\frac{E_{C M}^{2}}{2 m_{0}^{2}} =\frac{1 }{r\Delta _{r} }(2(r+\rho )a^2+2(r-\rho )r^2\\
			-(r-2\rho )L_1L_2-2a\rho (L_1+L_2)\\
			-\sqrt{2\rho(a-L_1)^2 +2\rho r^2-rL_1^2}\\ \sqrt{2\rho(a-L_2)^2 +2\rho r^2-rL_2^2}) ,
		\end{aligned}
	\end{array}
\end{equation}

here $2\rho=\frac{2 M r^{3}}{\left(r^{2}+g^{2}\right)^{\frac{3}{2}}}-\alpha \ln \frac{r}{|\alpha|}$.  As for the parameters of black holes $\alpha$ and $g$ in the above equation, we can discuss that for the rotating Bardeen black hole surrounded by perfect fluid dark matter, when the dark matter parameter $\alpha$ is set to zero, the black hole reverts to a rotating Bardeen black hole. Therefore, if $2\rho$ is replaced by  $2m=2\frac{Mr^3}{(r^2+g^2)^\frac{3}{2} }$ in Eq.(\ref{eq:Ecm}), the expression of the CM energy is consistent with that Ref.\cite{Ghosh:2015pra}. When the dark matter parameter $\alpha$ and magnetic charge $g$ are zero, the black hole reverts to a Kerr black hole. If $2\rho$ is replaced by 2 in Eq.(\ref{eq:Ecm}), the expression of the CM energy is consistent with that in Ref. \cite{Banados:2009pr}.

From the above equation, we can draw that there are two major types of parameters affecting the CM energy of the two particles. The first type of parameters are those related to the nature of the black hole itself, namely $a$, $g$ and $\alpha$; the second type of parameter is related to the physical quantity of the particle, namely $L$ and $m_0$. Of course, the CM energy is also affected by the radial coordinates of the two particles. We can adjust the parameters in Eq.(\ref{eq:Ecm}) to make the CM energy of the two particles diverge at the event horizon of the black hole. Nonetheless, the former type of parameter is the property of the black hole itself and we cannot change it, so we can consider the second type of parameter. But in the latter parameter, adjusting the mass of the particle obviously cannot make the CM energy diverge, so we can adjust the angular momentum of the two particles. Thus, we are inspired to discuss whether it is possible to adjust the angular momentum of two particles to make the event horizon of the extremal rotating Bardeen black hole surrounded by perfect fluid dark matter or the non-extremal rotating Bardeen black hole surrounded by perfect fluid dark matter be arbitrarily high at the CM energy.

For extreme black holes, the inner and outer event horizons coincide. Fig.~\ref{fig:Ecm1} shows the $E_{MC}$ vs $r$ of extreme black holes for different parameters $g$ and $\alpha$. From the figure, we can get a conclusion that when one of the two particles of the incident particle has critical angular momentum, the CM energy diverges near the event horizon of the black hole, while the particle with $L<L_c$ only contributes limited $E_{MC}$, and the particle with $L>L_c$ cannot reach the event horizon of the black hole.

For non-extreme black holes, the inner and outer event horizons are separated, and Fig.~\ref{fig:Ecm2} shows the $E_{cm}$ vs $r$ for non-extreme black holes for different parameters $g$, $a$ and $\alpha$. $E_{cm}$ diverges at the event horizon under the condition that the angular momentum of the particle satisfies a range determined by Eq.(\ref{eq:L}) and the angular momentum of a particle is the critical angular momentum. But at this time, it is found that the critical angular momentum is larger than the range satisfied by the angular momentum, so the $E_{cm}$ does not diverge. For example, when $g$ =0.3, $\alpha$=-0.005 and $a$=0.7, the range of angular momentum is -4.65933$\sim $3.05211, and the critical angular momentum is 4.50116. The latter value is larger than the range of angular momentum, so the CM energy does not diverge.

\section{CONCLUSION}
\label{sec:5}
In this paper, we studied the event horizon of a rotating Bardeen black hole surrounded by perfect fluid dark matter, and analyzed its possibility as a particle accelerator by studying the CM energy of two particles falling freely from infinity. The event horizon structure of the rotating Bardeen black hole surrounded by perfect fluid dark matter is more complex than that of the Kerr black hole and the rotating Bardeen black hole.
We found that when $g$ and $\alpha$ were determined, one can get the critical value $a_E$, which corresponds to the extreme value of a degenerate horizon black hole, that is, when $a=a_E$, the two horizons coincide, and when $a<a_E$, the black hole is a regular black hole with Cauchy and event horizons, and when $a>a_E$, the black hole horizon does not exist.

We used the BSW mechanism to obtain the expression of the CM energy of two particles colliding in the equatorial plane of the black hole and analyze its properties. Through the conservation of energy and angular momentum of the particle and the four-speed normalization of the particle, the equation of motion of the particle was obtained, and the value range of angular momentum of the particle was obtained. We calculated the expression of particle CM energy and discovered that for extreme black holes, when one of the incident particles has critical angular momentum, its CM energy will diverge. Nevertheless, in the case of non-extreme black holes, the CM energy is limited, because the particle meeting the critical angular momentum cannot reach the event horizon of the black hole, and its angular momentum does not satisfy the value range of the angular momentum of the particle.  For particles that can be incident into a non-extreme black hole, the CM energy depends on the magnetic charge parameter $g$ and the dark matter parameter $\alpha$, and the rotation parameter $a$, so the BSW mechanism depends on those parameters as well.

\end{document}